\documentclass[11pt,a4paper]{article}
\usepackage{amsmath,amssymb,bm,epsfig,color,graphicx,cite,braket,mathrsfs,slashed,multirow,hyperref}
\newcommand{\sng}{SNOwGLoBES}
\newcommand{\lib}[1]{\texttt{#1}}
\newcommand{\eg}{\textit{e.g.}}

\renewcommand{\thefootnote}{\fnsymbol{footnote}}

\usepackage{hyperref,xcolor,doi}
\hypersetup{
    colorlinks=true,
    linktocpage=true,
    linkcolor=blue,
    filecolor=blue,      
    urlcolor=blue,
    citecolor=blue,
    }

\begin{document}

\title{
\begin{flushright}
\begin{minipage}{0.2\linewidth}
\normalsize
CTPU-PTC-22-13 \\*[50pt]
\end{minipage}
\end{flushright}
{\Large \bf 
Probing non-standard neutrino interactions with a light boson\\ from next galactic and diffuse supernova neutrinos
\\*[20pt]}}

\author{
Kensuke Akita\footnote{
E-mail: \href{mailto:kensuke8a1@ibs.re.kr}{kensuke8a1@ibs.re.kr}}
,\ \  Sang Hui Im\footnote{E-mail:  \href{mailto:imsanghui@ibs.re.kr}{imsanghui@ibs.re.kr}}
\ \ and\ \ Mehedi Masud\footnote{E-mail: \href{mailto:masud@ibs.re.kr}{masud@ibs.re.kr}}\\*[20pt]
{\it
Center for Theoretical Physics of the Universe,}\\
{\it
Institute for Basic Science,
Daejeon 34126, Korea} \\*[50pt]
}

\date{
\centerline{\small \bf Abstract}
\begin{minipage}{0.9\linewidth}
\medskip \medskip \small 
Non-standard neutrino interactions with a massive boson can produce the bosons in the core of core-collapse supernovae (SNe). After the emission of the bosons from the SN core, their subsequent decays into neutrinos can modify the SN neutrino flux. We show future observations of neutrinos from a next galactic SN in Super-Kamiokande (SK) and Hyper-Kamiokande (HK) can probe flavor-universal non-standard neutrino couplings to a light boson, improving the previous limit from the SN 1987A neutrino burst by several orders of magnitude.
We also discuss sensitivity of the flavor-universal non-standard neutrino interactions in future observations of diffuse neutrinos from all the past SNe, known as the diffuse supernova neutrino background (DSNB). According to our analysis, observations of the DSNB in HK, JUNO and DUNE experiments can probe such couplings by a factor of $\sim 2$ beyond the SN 1987A constraint. However, our result is also subject to a large uncertainty concerning the precise estimation of the DSNB.
\end{minipage}
}

\maketitle{}
\thispagestyle{empty}
\clearpage
\tableofcontents
\clearpage

\renewcommand{\thefootnote}{\arabic{footnote}}
\setcounter{footnote}{0}

\section{Introduction}
\label{sec1}
Neutrinos are still mysterious particles. The observations of neutrino oscillations have established the fact that at least two neutrino masses are non-zero but much smaller than the electron mass.
The lack of an explanation for such small neutrino masses in the SM suggests the existence of non-standard neutrino interactions with a new particle. 
The most promising classes of models to explain the origin of the small neutrino masses would be those that involve the seesaw mechanism \cite{Minkowski:1977sc, Mohapatra:1979ia, Gell-Mann:1979vob, Yanagida:1979as,Schechter:1980gr}. In the seesaw mechanism, heavy right-handed neutrinos with a Majorana mass term naturally suppress the masses of left-handed neutrinos. A large Majorana mass term in the seesaw mechanism requires that the active neutrinos are Majorana particles, and the lepton number symmetry is broken.
Assuming that the lepton number symmetry is actually a global symmetry of more fundamental theory as in the SM and becomes spontaneously broken at low energies,
there is an associated pseudo-Nambu-Goldstone boson, called the Majoron \cite{Chikashige:1980ui,Gelmini:1980re,Schechter:1981cv}. The Majoron has to couple to neutrinos.

A large non-standard neutrino interaction with a light particle can potentially explain short baseline neutrino anomalies \cite{Asaadi:2017bhx,Chauhan:2018dkd,Smirnov:2021zgn,Dentler:2019dhz,deGouvea:2019qre,Jeong:2018yts, Abdallah:2022grs}, muon $g-2$ anomaly \cite{Muong-2:2006rrc,Araki:2015mya,Borsanyi:2020mff,Muong-2:2021ojo,Carpio:2021jhu} and tensions in cosmology \cite{vandenAarssen:2012vpm, Cyr-Racine:2013jua,Cherry:2014xra,Chu:2015ipa,Lancaster:2017ksf,Chu:2018gxk,Kreisch:2019yzn,Grohs:2020xxd, Das:2020xke}.
Relatively smaller neutrino interactions with a new light gauge boson could also help to resolve the Hubble tension \cite{Escudero:2019gzq}.
Moreover, non-standard neutrino interactions with a new light boson would have considerable effects on cosmology and particle experiments. Such scenarios are constrained by the observations of the Cosmic Microwave Background (CMB) \cite{Archidiacono:2013dua,Forastieri:2015paa,Escudero:2019gvw, Forastieri:2019cuf,Blinov:2019gcj,RoyChoudhury:2020dmd,Brinckmann:2020bcn,Venzor:2022hql}, Big Bang Nucleosynthesis (BBN) \cite{Saviano:2014esa,Huang:2017egl,Escudero:2019gvw, Venzor:2020ova}, high-energy astrophysical neutrinos \cite{Ioka:2014kca,Ibe:2014pja,Shoemaker:2015qul,Bustamante:2020mep,Esteban:2021tub,Carpio:2022lqk} and other experiments of particle decays \cite{Laha:2013xua,Berryman:2018ogk,Brune:2018sab}. 

Core-collapse supernovae (SNe) also provide means to test non-standard neutrino interactions with a light boson.
SN is a powerful source of neutrinos, where in each explosion $\mathcal{O}(10^{58})$ neutrinos and anti-neutrinos of all flavors are emitted on a time scale of $t\sim 10\ {\rm s}$ with average energies $E\sim 15\ {\rm MeV}$ (see e.g., Refs.~\cite{Mirizzi:2015eza,Horiuchi:2018ofe} for recent reviews).
Production of light bosons via feeble non-standard neutrino interactions can induce a channel for the energy loss of the SN core \cite{Raffelt:2012kt}. Such interactions are constrained by the observation of SN 1987A neutrino burst \cite{Brune:2018sab, Heurtier:2016otg}.
In addition, considerations of neutrino self-interactions mediated by a new light boson in the SN core \cite{ Yang:2018yvk,Shalgar:2019rqe,Babu:2019iml, Das:2017iuj,Cerdeno:2021cdz} as well as the scattering between SN neutrinos and the Cosmic Neutrino Background (C$\nu$B) \cite{Kolb:1987qy,Shalgar:2019rqe,Das:2022xsz} impose constraints on large non-standard neutrino interactions by the SN 1987A data \cite{Kolb:1987qy,Shalgar:2019rqe} and the Super-Kamiokande data \cite{Das:2022xsz}. 
Such large neutrino interactions with a light boson can also alter the dynamics of the core-collapse and can be constrained \cite{Fuller:1988ega,Chang:2022aas}.\footnote{For \eg, \cite{Manohar:1987ec} argues that neutrinos with self interactions are prevented from escaping the protoneutron star (PNS) resulting in the modification of neutrino signal duration by the neutrino diffusion time inside the PNS. On the other hand,  the authors of \cite{Dicus:1988jh} claim that no limit on the interaction can be obtained on the basis of the expansion of a tightly-coupled neutrino fluid irrespective of the strength of neutrino self-interaction. Recently, \cite{Chang:2022aas} shows that for a sufficient strength of interaction, the neutrino signal duration can be significantly modified for a burst-outflow of neutrino from a tightly interacting neutrino fluid. In our analysis, the coupling strength small enough so that we safely ignore the modifications of the core-collapse dynamics.}
The supernova constraints on a massless or very light boson coupled to neutrinos have also extensively been studied over the past few decades \cite{Gelmini:1982rr, Kolb:1987qy,Choi:1987sd,Konoplich:1988mj,BEREZHIANI1989279,Choi:1989hi,Chang:1993yp,Kachelriess:2000qc,Tomas:2001dh,Lindner:2001th,Hannestad:2002ff,Farzan:2002wx,Ando:2003ie,Fogli:2004gy}.

In this work we discuss the future sensitivity on probing flavor-universal neutrino interactions with a new light boson from the observation of SNe neutrinos.
Light bosons produced via such interactions in the SN core would decay into neutrinos in the stellar envelope.
With a sufficiently long lifetime of the light boson, the light boson decays into neutrinos in a region beyond the high stellar density region, and the produced neutrinos are not thermalized. 
It can give rise to a significant modification of the spectrum of the SN neutrino flux.
We show that even when such interactions are too small to yield a significant energy loss in the core, observable modifications of neutrino flux can be made by the decay process.
Alternative scenarios for modifications of the neutrino flux and $\gamma$-rays from SNe due to decays of exotic particles, e.g., axion-like particles \cite{Calore:2020tjw,Ferreira:2022xlw}, dark photons \cite{DeRocco:2019njg,Calore:2021lih} and heavy sterile neutrinos \cite{Calore:2021lih,Mastrototaro:2019vug} have been studied.

We focus on two SN neutrino sources: (i) a next future galactic SN, (ii) all the past core-collapse supernovae in the universe. If a next galactic SN event occurs, a large number of neutrinos will be observed at the large neutrino detectors such as Super-Kamiokande (SK) and Hyper-Kamiokande (HK) \cite{Hyper-Kamiokande:2018ofw}. However, the SN events are rare, happening a few times per century \cite{Rozwadowska:2020nab}.
On the other hand, diffuse neutrino flux coming from all the past SNe, dubbed the diffuse supernova neutrino background (DSNB) (see e.g., Refs.~\cite{Mirizzi:2015eza,Beacom:2010kk,Lunardini:2010ab} for recent reviews), is a stationary and predictable flux. Although the DSNB have not yet been detected (see Ref.~\cite{Super-Kamiokande:2021jaq} for the latest upper bound of the DSNB flux in SK), large underground detectors such as HK, JUNO \cite{JUNO:2015zny} and DUNE \cite{DUNE:2020ypp} will be capable of probing the DSNB with high statistics.

This article is organized as follows.
In Section~\ref{sec2}, we introduce non-standard neutrino interactions with a new boson. In Section~\ref{sec3}, we discuss the energy spectrum of a light boson emitted in the SN core. In Section~\ref{sec4}, we discuss the flux of neutrinos produced by the boson decays.
In Sections~\ref{sec5} and \ref{sec6}, we estimate the discovery potentials for next galactic supernova neutrinos and diffuse supernova neutrinos produced by the scalar boson decays, respectively. In Section~\ref{sec7}, we give our conclusions.
In appendix~\ref{appa}, we estimate the discovery potentials for supernova neutrinos from vector boson decays. The review of constraints on non-standard neutrino interactions with a massive scalar boson form the energy loss of SN 1987A is discussed in appendix~\ref{appb}.

\section{Non-standard neutrino interactions with a light boson}
\label{sec2}

We will consider neutrino interactions with a new scalar boson, which may be described by
\begin{align}
    \mathcal{L}=\frac{1}{2}g_{\alpha\beta}\bar{\nu}_\alpha\nu_\beta\phi,
    \label{NNI}
\end{align}
where $\nu$ is a 4-component field, and $\alpha,\beta=e,\mu,\tau$.
For Majorana neutrinos, $\nu$ is composed of left-handed active neutrinos and right-handed active neutrinos (i.e., right-handed active anti-neutrinos). For Dirac neutrinos, $\nu$ is composed of left-handed active neutrinos and right-handed sterile neutrinos.
For simplicity, here we consider only the scalar interaction.  
Neutrino interactions with a pseudo-scalar boson give the same results when neutrinos are ultra-relativistic as the neutrinos emitted from SNe. 
Therefore the following analysis is applicable to the case of pseudo-scalar bosons.
Interactions with vector bosons also give similar results up to  $\mathcal{O}(1)$ factor difference for the squared matrix amplitudes when neutrinos are ultra-relativistic but the production processes of vector bosons in the SN core are different from the case of the scalar bosons. We will discuss the case of vetcor bosons in appendix~\ref{appa}. 

For simplicity, we will also assume that neutrinos are Majorana particles and the coupling matrix $g_{\alpha \beta}$ is universal and diagonal, i.e.
\begin{align}
    g_{ee}=g_{\mu\mu}=g_{\tau\tau}=g.
    \label{NNI2}
\end{align}
The dominant production processes for $\phi$ in the SN core are $\nu_\alpha\nu_\alpha\rightarrow \phi$ and $\bar{\nu}_\alpha\bar{\nu}_\alpha\rightarrow \phi$. Subsequently the decay processes $\phi\rightarrow \nu_\alpha\nu_\alpha$ and $\phi\rightarrow \bar{\nu}_\alpha\bar{\nu}_\alpha$ occur for a scalar boson heavier than two neutrinos.
Here $\nu_\alpha$ and $\bar{\nu}_\alpha$ denote left-handed neutrinos and right-handed (anti-)neutrinos, respectively, not the 4-component field.
In particular, $\nu_e\nu_e\rightarrow \phi$ is the most dominant production process, because in the SN core  $\nu_e$ are dominantly produced by electron capture processes compared with the other neutrinos and anti-neutrinos.
The squared matrix amplitude for $\phi\rightarrow \nu_\alpha \nu_\alpha$ and the decay rate in the rest frame of $\phi$ are
\begin{align}
    |\mathcal{M}|_{\phi\rightarrow \nu_\alpha\nu_\alpha}^2&=g^2m_\phi^2, \\
    \Gamma_{\phi\rightarrow \nu_\alpha\nu_\alpha}
    &=\frac{g^2}{32\pi}m_\phi.
\end{align}
The same expressions hold for the processes $\phi\rightarrow \bar{\nu}_\alpha \bar{\nu}_\alpha$.
On the other hand, for the case of interactions with a vector boson $Z'$, the dominant production processes for $Z'$ in the SN core are $\nu_\alpha\bar{\nu}_\alpha\rightarrow Z'$. In particular, the production rates for $\nu_e\bar{\nu}_e\rightarrow Z'$ in the core are typically much smaller than that for $\nu_e\nu_e\rightarrow \phi$ since $\nu_e$ are dominantly produced by electron capture process in the SN core.

We comment the case of Dirac neutrinos. For Dirac neutrinos, we can consider two types of the interaction Lagrangians with a scalar boson \cite{Chang:2022aas},
\begin{align}
\mathcal{L}^1&=g_{\alpha\beta}\overline{\nu^c_\alpha}\nu_\beta\phi + {\rm h.c.}, \\
\mathcal{L}^2&=g_{\alpha\beta}\bar{\nu}_\alpha\nu_\beta\phi,
\end{align}
where $\nu^c$ is the charge conjugate of the neutrino. The light boson has the lepton number $L_\phi=-2$ and $L_\phi=0$ for the first and second case, respectively. 
For the first type of Lagrangian, the dominant production process of $\phi$ is $\nu\nu\rightarrow \phi^\ast$ and $\bar{\nu}\bar{\nu}\rightarrow \phi$.
Thus, for the first Lagrangian of the Dirac neutrinos, the following results for the Majorana neutrinos can be applied up to $\mathrm{O}(1)$ factors.
On the other hand, for the second Lagrangian of the Dirac neutrinos, the production process of $\phi$ is $\bar{\nu}\nu \rightarrow \phi$. Since right-handed neutrinos and left-handed antineutrinos are not produced via the Standard Model (SM) interactions in the SN core, the productions of light bosons would also be suppressed unless right-handed neutrinos and left-handed anti-neutrinos are populated due to helicity flipping of large secret neutrino self-interactions mediated by light bosons \cite{Esteban:2021tub,Chang:2022aas}. We leave the detailed estimation of the second Lagrangian of Dirac neutrinos as future work. In the following we consider the case of Majorana neutrinos described in Eqs.~(\ref{NNI}) and (\ref{NNI2}) as mentioned above.

\section{Light boson emission from supernovae}
\label{sec3}
In this section we describe the emission of a light scalar boson $\phi$ by the Boltzmann equations for the system of $\nu$ and $\phi$ in the SN core. We then compute the energy spectrum of the scalar boson $\phi$ emitted from the SN core.

As a reference model for a supernova, we consider an isotropic and homogeneous supernova with the temperature inside the core of $T\sim30\ {\rm MeV}$ and the core radius of $r_c\sim10\ {\rm km}$ on a time scale of $\Delta t\sim 10\ {\rm s}$.
In the SN core, the electron neutrinos $\nu_e$ are much more than the anti-electron neutrinos $\bar{\nu}_e$ due to electron capture processes.
Thus the chemical potential $\mu_{\nu_e}$ for $\nu_e$ is a large positive value, whereas the chemical potential $\mu_{\bar{\nu}_e}$ for $\bar{\nu}_e$ is the opposite negative value. 
We assume $\mu_{\nu_e}\sim 200\ {\rm MeV}$ and $ \mu_{\bar{\nu}_e}\sim -200\ {\rm MeV}$ during $\Delta t$ , which are the corresponding Fermi energy for $\nu_e$ in the core \cite{Burrows:1986me}.
On the other hand, since $\nu_\mu, \nu_\tau$ and their anti-particles have the same interactions in a first-order approximation, their chemical potentials are approximately zero.
We leave precise numerical simulations of supernovae to estimate the precise values of these parameters as future work.
For a more accurate treatment of the SN core in physics beyond the SM, see e.g., Refs.~\cite{Mastrototaro:2019vug,Balaji:2022noj}.
Non-standard neutrino interactions with a light boson can alter the dynamics of the core-collapse if the magnitude of such interactions are enough strong \cite{Chang:2022aas}. For smaller magnitude of such interactions of our interest, the dynamics of the core-collapse cannot be altered.

For the production processes $\nu_\alpha(p_1)\nu_\alpha(p_2)\rightarrow \phi(p_\phi)$ and $\bar{\nu}_\alpha(p_1)\bar{\nu}_\alpha(p_2)\rightarrow \phi(p_\phi)$, the emission rate of the number of $\phi$ per unit volume of the SN core and per unit time is given by 
\begin{align}
    \frac{dn_\phi}{dt}=\sum_\nu\int d\Pi_\phi d\Pi_1d\Pi_2 S|\mathcal{M}|^2_{\nu\nu\rightarrow\phi}(2\pi)^4\delta^4(p_1+p_2-p_\phi)f_\nu(p_1)f_\nu(p_2),
    \label{Bphi}
\end{align}
where $\nu=\nu_\alpha$ and $\bar{\nu}_\alpha\ (\alpha=e,\mu,\tau)$, $d\Pi=d^3p/[(2\pi)^32E]$ is the phase space, and $S=1/2$ is the symmetry factor for identical particles in the initial state. Since neutrinos are in the thermal equilibrium in the SN core, the distributions for neutrinos are given by the Fermi-Dirac distribution with the chemical potential $\mu_\nu$,
\begin{align}
    f_\nu(p)=\frac{1}{e^{(p-\mu_\nu)/T}+1}.
\end{align}
where $\mu_{\nu_e}\sim 200\ {\rm MeV},\ \mu_{\bar{\nu}_e}\sim -200\ {\rm MeV}$ and the other chemical potentials are zero \cite{Burrows:1986me}. 

If the produced light bosons decay into neutrinos inside the neutrino-sphere where neutrinos are trapped through their interactions with nucleons of the medium, the observed neutrinos from a supernova would be the same regardless of the existence of the light bosons.
To take into account this effect, we plug the decay factor $e^{-\Gamma_\phi r_\nu/\gamma}$ into Eq.~(\ref{Bphi}), where we assume $r_\nu\sim 50\ {\rm km}$
\footnote{The radius of neutrino-sphere depends on neutrino energy. Since neutrinos produced by the light boson decay are more energetic in the stellar envelope, their radius of neutrino-sphere would be larger than the typical radius for the standard SN neutrinos. We leave detailed estimation of this radius as future work. However, the radius would not change dramatically, because the density of the medium in the stellar envelope decreases rapidly with $\rho \propto r^{-3}$ \cite{1989A&AS...78..375J}.}.
$\Gamma_\phi$ is the total decay rate of $\phi$ in the rest frame of $\phi$, and $\gamma=E_\phi/m_\phi$ is the boost factor. After straightforward calculations, we obtain
\begin{align}
     \frac{dn_\phi}{dt}=\sum_\nu\frac{g^2 m_\phi^2}{64\pi^3}\int_0^\infty \frac{p_\phi dp_\phi}{E_\phi}e^{-\Gamma_\phi r_\nu/\gamma} \int_{p_1^{\rm min}}^{p_1^{\rm max}} dp_1\ f_\nu(p_1)f_\nu(E_\phi-p_1),
\end{align}
where $p_1^{\rm min}=m_\phi^2/[2(E_\phi+p_\phi)]$ and $p_1^{\rm max}=m_\phi^2/[2(E_\phi-p_\phi)]$.
Integrating over the SN core in 10 second interval one obtains the total spectrum for the light boson from a supernova,
\begin{align}
    \frac{dN_\phi}{dE_\phi}=V\Delta t \sum_\nu\frac{g^2 m_\phi^2}{64\pi^3}e^{-\Gamma_\phi r_\nu/\gamma}\int_{p_1^{\rm min}}^{p_1^{\rm max}} dp_1\ f_\nu(p_1)f_\nu(E_\phi-p_1),
    \label{phispectrum}
\end{align}
where $V=\frac{4\pi}{3}r_c^3$ and $\Delta t=10\ {\rm s}$.

Here we comment on the matter effect on neutrinos in the SN core. The weak interactions of neutrinos with a thermal background produce the effective neutrino masses, corresponding to the effective potentials of neutrinos. In the SN core, the typical values of effective potentials for $\overset{(-)}{\nu}_e$ and $\overset{(-)}{\nu}_{\mu,\tau}$ are of the order of $|V_e| \sim 1\ {\rm eV}$ and $|V_{\mu,\tau}|\sim 10\ {\rm eV}$, respectively.
They modify the decay rate as $\Gamma_{\phi}/\gamma\sim g^2(m_\phi^2/E_\phi-2V_\alpha)$ \cite{Brune:2018sab}. Thus for $m_{\phi}\gtrsim 10\ {\rm keV}$, the effective potentials are negligible. For $m_{\phi}\lesssim 10\ {\rm keV}$, the effective potentials can be important. Yet the CMB and BBN observations impose severe constraints for $m_{\phi}\lesssim 10\ {\rm keV}$ \cite{Archidiacono:2013dua,Escudero:2019gvw, Huang:2017egl}. Therefore we can neglect the effective potentials of neutrinos.

We also make a comment on the trapping of $\phi$ in the core by the scattering processes $\overset{(-)}{\nu}\phi \rightarrow \overset{(-)}{\nu} \phi$ \cite{Brune:2018sab, Heurtier:2016otg}. If the interactions of $\phi$ with neutrinos are large enough, the light bosons are trapped in the SN core, modifying the emitted spectrum of $\phi$. The precise quantitative estimation is quite challenging. However, such large interactions with $m_{\phi} \lesssim 100\ {\rm keV}$ are mostly excluded by the CMB \cite{Archidiacono:2013dua,Escudero:2019gvw} and BBN \cite{Escudero:2019gvw,Huang:2017egl}. On the other hand, for $m_{\phi} \gtrsim 100\ {\rm keV}$, such large interactions will make the rapid decays of $\phi$, not producing the spectrum of $\phi$ outside the neutrino-sphere. Therefore we do not consider this effect.

Finally the gravitational potential of the supernova could trap the light bosons inside the core, preventing their emissions. To determine the mass of $\phi$ where the light bosons are trapped inside the core, we follow the simple argument introduced in Ref.~\cite{Dreiner:2003wh}.
The gravitational trapping effect occurs when the kinetic energy of the light boson satisfies
\begin{align}
    E_{\rm kin}\leq K_{\rm tr}= \frac{G_NM_cm_\phi}{r_c},
    \label{Potentialenergy}
\end{align}
where $G_N$ is the Newton constant, $M_c\sim M_\odot$ is the enclosed mass of the supernova inside the core radius $r_c$.
One can take into account this effect by modifying the energy spectrum for the scalar boson \cite{Mastrototaro:2019vug},
\begin{align}
    \frac{dN_\phi}{dE_\phi} \,\rightarrow\, \frac{dN_\phi}{dE_\phi}\,\chi(E_\phi-K_{\rm tr}-m_\phi),
\end{align}
where $\chi$ is the Heaviside step function.

\begin{figure}
	\begin{center}
	\includegraphics[clip,width=10cm]{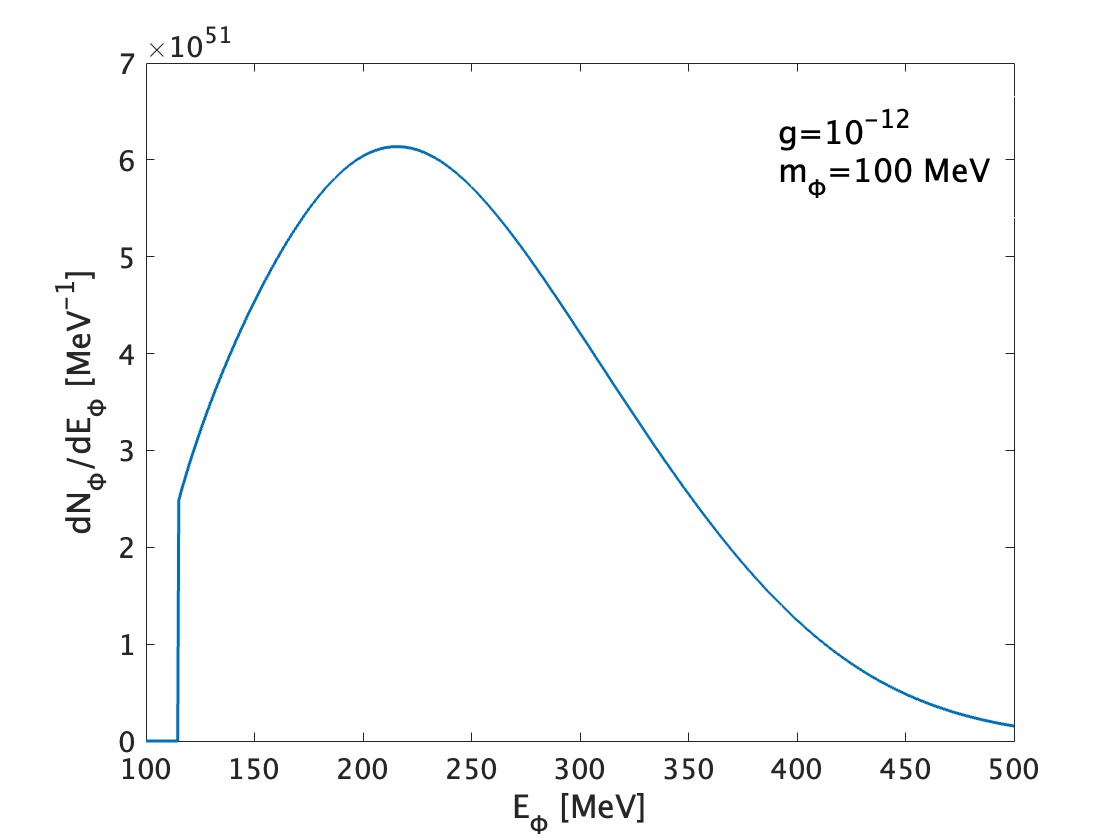}
	\end{center}
	 \vspace{-6mm}
	\caption{The energy spectrum of the scalar boson emitted from a supernova with $g=10^{-12}$ and $m_{\phi}=100\ {\rm MeV}$.}
	\label{fig:Majoron_spectrum}
\end{figure}

In Fig.~\ref{fig:Majoron_spectrum}, we show the spectrum for the scalar boson in the case of $g=10^{-12}$ and $m_{\phi}=100\ {\rm MeV}$. The peak of the spectrum appear around $E_{\phi}\sim 200\ {\rm MeV}$ because the electron neutrinos in the core have a typical energy of $200\ {\rm MeV}$ due to $\mu_{\nu_e}\sim 200\ {\rm MeV}$, inducing the dominant production process of $\nu_e\nu_e\rightarrow \phi$. The cutoff at $E_\phi \simeq 120\ {\rm MeV}$ is due to the gravitational trapping. 
For a larger mass of $\phi$, the cutoff scale from the gravitational trapping is larger, and it has a substantial effect on the energy spectrum for $m_{\phi}\gtrsim 300\ {\rm MeV}$.

\section{Neutrino fluxes from boson decays}
\label{sec4}
The light bosons decaying outside the neutrino-sphere may produce an observable effect on the neutrino fluxes from supernovae. In this section we estimate the modification of the neutrino fluxes from a next future galactic supernova and from all the past supernovae in the universe.

\subsection{Flux from a next galactic supernova}
First we discuss a flux of neutrinos from a next future galactic supernova, following Refs.~\cite{Mastrototaro:2019vug,Oberauer:1993yr}.
Suppose that the light boson decays at a time $t_D$ and at a distance $R_D=\beta t_D$ from the SN with its velocity $\beta=v_\phi/c$.
Neutrinos will be emitted at an angle $\Theta$ relative to the light boson momentum, where we define $\cos\Theta\equiv \bm{p}_\phi\cdot \bm{p}_\nu/\left(|\bm{p}_\phi||\bm{p}_\nu|\right)$. After the light boson decays, the produced neutrinos will travel a distance $\ell_\nu$ until they arrive at the Earth.
The distance between the center of the SN and the Earth is $d_{\rm SN}$. 
Then $\ell_\nu$ can be written, using the law of cosine and assuming $R_D \ll d_{\rm SN}$, which is valid for parameter region of our interest
\begin{align}
    d_{\rm SN}^2&=\ell_{\nu}^2+R_D^2+2\ell_\nu R_D\cos\Theta, \nonumber \\
    \ell_\nu&\simeq d_{\rm SN}-R_D\cos\Theta.
    \label{distance}
\end{align}
The arrival time for the emitted neutrinos is $t_{\rm arr}=t_D+\ell_\nu/c$ while the minimum arrival time is $t_{\rm min}=d_{\rm SN}/c$. Then the delayed time $t$ for arrival is given by
\begin{align}
    t&=t_{\rm arr}-t_{\rm min}, \nonumber \\
    &=t_D+\ell_\nu-d_{\rm SN}, \nonumber \\
    &=t_D(1-\beta\cos\Theta),
\end{align}
where we set $c=1$ and use Eq.~(\ref{distance}) and $R_D=\beta t_D$.
We transform the angle $\Theta$ to another one $\theta$, which is the angle of neutrino emission in the rest frame of $\phi$ relative to $\bm{p}_\phi$ in the lab frame:
\begin{align}
    \cos\Theta=\frac{\beta+\cos\theta}{1+\beta\cos\theta}.
\end{align}
We thus obtain
\begin{align}
    t=\frac{t_D}{\gamma^2(1+\beta\cos\theta)},
\end{align}
where $\gamma=1/\sqrt{1-\beta^2}=E_\phi/m_\phi$.
Then the number of $\nu_\alpha$ arriving at a spherical shell with a radius of $d_{\rm SN}$ per $t_D$ and $\omega_\nu$, where $\omega_\nu$ is the energy of the emitted neutrinos in the rest frame of $\phi$, is given by \cite{Oberauer:1993yr}
\begin{align}
\frac{dN_{\nu_\alpha}}{dt_Dd\omega_\nu}&=B_\alpha \bar{N}_\nu \int d\cos\theta \int dE_\phi \frac{1}{\tau\gamma} \frac{dN_\phi(t_D, E_\phi)}{dE_\phi}f_{\nu_\alpha}(\omega_\nu,\cos\theta), \nonumber \\
&=B_\alpha \bar{N}_\nu\int d\cos\theta \int dE_\phi \frac{1}{\tau\gamma} \exp\left(-\frac{t_D}{\tau\gamma} \right)\frac{dN_\phi(0, E_\phi)}{dE_\phi}f_{\nu_\alpha}(\omega_\nu,\cos\theta),
\end{align}
where $\tau$ is the lifetime of $\phi$ in the rest frame, $B_\alpha$ is the branching ratio of the considered decay process and $\bar{N_\nu}$ is the number of the emitted neutrinos by the decay of a particle $\phi$. For $\phi\rightarrow\nu_\alpha\nu_\alpha$, $\bar{N}_\nu=2$.
$B_\alpha\bar{N}_\nu/(\tau\gamma)\exp(-t_D/(\tau\gamma))$ is the production rate for neutrinos per time, i.e. $dN_\phi/dt_D$.
The spectrum $dN_\phi(0,E_\phi)/dE_\phi$ is given in Eq.~(\ref{phispectrum}). Finally $f_{\nu_\alpha}(\omega_\nu,\cos\theta)$ is a distribution function for $\omega_\nu$ and $\theta$, which is normalized to be 1.

Let us change variables to the delayed time and the neutrino energy in the lab frame by the usual transformation $E_{\nu}=\gamma(1+\beta\cos\theta)\omega_\nu$. We obtain a differential flux as
\begin{align}
    &\frac{dN_{\nu_\alpha}}{dtdE_\nu} \nonumber \\
    &=\frac{B_\alpha\bar{N}_\nu}{\tau}\int d\cos\theta \int dE_\phi \exp\left[-\frac{\gamma(1+\beta\cos\theta)t}{\tau} \right]\frac{dN_\phi(0, E_\phi)}{dE_\phi}f_{\nu_\alpha}\left(\frac{E_\nu}{\gamma(1+\beta\cos\theta)},\cos\theta\right).
\end{align}
Integrating over the delayed time, the expression for the neutrino energy spectrum is obtained as
\begin{align}
    \frac{dN_{\nu_\alpha}}{dE_\nu}=B_\alpha\bar{N}_\nu
    \int d\cos\theta \int dE_\phi
    \frac{1}{\gamma(1+\beta\cos\theta)}\frac{dN_\phi(0, E_\phi)}{dE_\phi}f_{\nu_\alpha}\left(\frac{E_\nu}{\gamma(1+\beta\cos\theta)},\cos\theta\right).
    \label{Flux_nu}
\end{align}
For $\phi\rightarrow \nu_\alpha\nu_\alpha$, the emitted neutrinos in the rest frame of $\phi$ are isotropic and have energy of $\omega_\nu=m_\phi/2$. Then the distribution function for $\omega_\nu$ and $\cos\theta$ is given by the delta function:
\begin{align}
    f_{\nu_\alpha}(\omega_\nu,\cos\theta)=\frac{1}{2}\delta\left(\omega_\nu-\frac{m_\phi}{2}\right).
    \label{DF}
\end{align}
Plugging Eq.~(\ref{DF}), $B_a=1/6$, and $\bar{N}_\nu=2$ into Eq.~(\ref{Flux_nu}), we obtain $dN_{\nu_\alpha}/dE_\nu$,
\begin{align}
    \frac{dN_{\nu_\alpha}}{dE_\nu}=\frac{1}{3}\int dE_\phi \frac{1}{p_\phi} \frac{dN_\phi(0,E_\phi)}{dE_\phi}.
    \label{FNfinal}
\end{align}
The integral range of $E_\phi$ is the region satisfying $m_\phi^2/[2(E_\phi+m_\phi)]\leq E_\nu\leq m_\phi^2/[2(E_\phi-m_\phi)$, which represents the kinematically allowed relationship between $E_\phi$ and $E_\nu$.

\begin{figure}
	\begin{center}
	\includegraphics[clip,width=10cm]{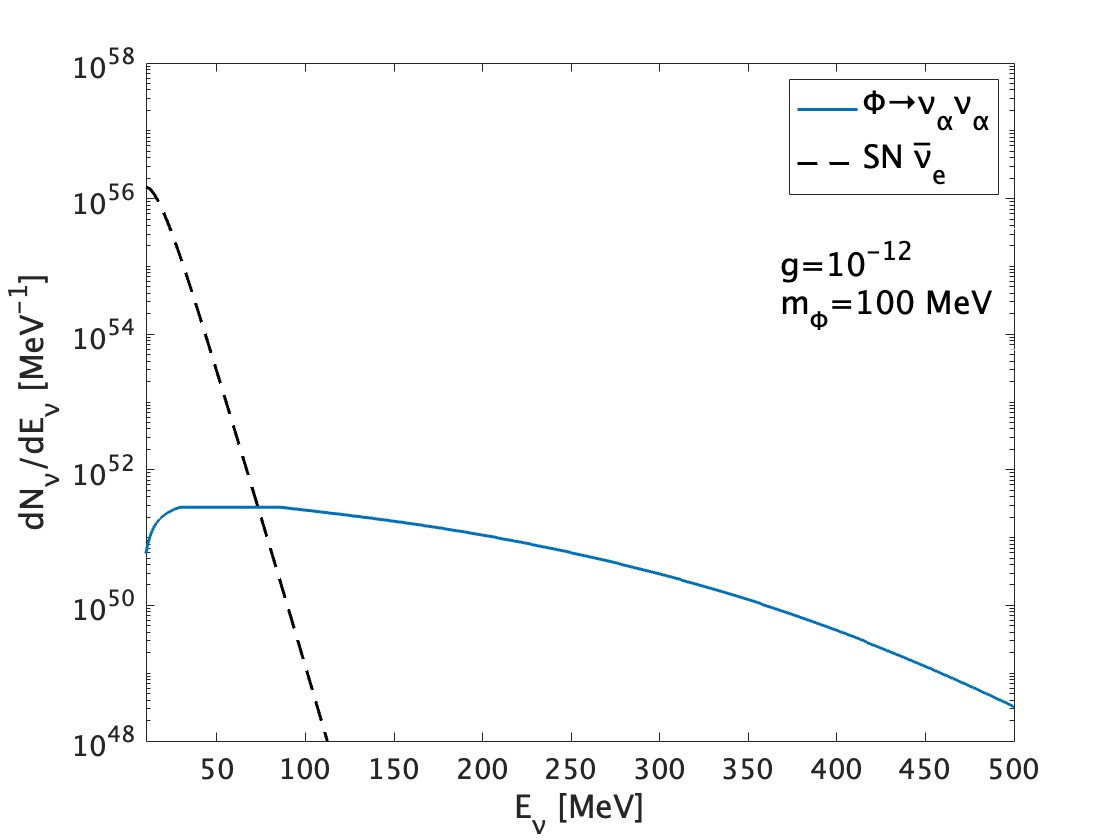}
	\end{center}
	 \vspace{-6mm}
	\caption{The energy distribution of neutrinos produced by the decays of the scalar bosons, $\phi\rightarrow \nu_\alpha\nu_\alpha$, emitted from a supernova with $g=10^{-12}$ and $m_{\phi}=100\ {\rm MeV}$ (blue solid line, Eq.~(\ref{FNfinal})) and anti-electron neutrinos from a supernova (black dashed line, Eq.~(\ref{SNnu})).}
	\label{fig:Nuspectrum_Mdecay}
\end{figure}

In Fig.~\ref{fig:Nuspectrum_Mdecay}, we show the energy distribution of neutrinos produced by the decays of the scalar bosons emitted from the SN core with $g=10^{-12}$ and $m_{\phi}=100\ {\rm MeV}$ (blue solid line), which is given by Eq.~(\ref{FNfinal}). We also show the distribution of neutrinos directly produced from SN explosion (black dashed line), whose details are discussed in Section \ref{sec5.1} and given by Eq.~(\ref{SNnu}). Neutrinos produced by the decays of $\phi$ is dominant at $E_\nu \gtrsim 80\ {\rm MeV}$.
Even for the small coupling of $g=10^{-12}$, the energy distribution of SN neutrinos is modified due to the large number density of neutrinos (mainly of electron neutrinos with $\mu_{\nu_e}\sim 200\ {\rm MeV}$).

Finally we make a comment on neutrino flavor conversions in SNe. Since we assume the flavor-universal coupling to a light boson $(g_{ee}=g_{\mu\mu}=g_{\tau\tau})$, flavor conversions of neutrinos after the light boson decay is already averaged out and ineffective. In general, however, flavor-dependent couplings can modify the resultant neutrino fluxes at the Earth by neutrino oscillations in SNe.  
In density region close to the neutrino sphere, the high neutrino density might affect the resultant neutrino flux by means of a refractive effect via neutrino self-interactions. But this effect is still under investigation (e.g., see Refs.~\cite{Duan:2010bg,Chakraborty:2016yeg,Tamborra:2020cul,Mirizzi:2015eza} for recent reviews). In the SN envelope at lager radii, the flavor conversions of neutrinos after the light boson decay might be enhanced by neutrino interactions with the electron background called the Mikheev-Smirnov-Wolfstein (MSW) effect \cite{Wolfenstein:1977ue,Mikheyev:1985zog,Mikheev:1986if}. Both effects would depend on the lifetime of the light boson and the emitted neutrino energy. We leave this analysis for flavor-dependent non-standard neutrino interactions as future study.

\subsection{Flux from all the past supernovae}
Next we discuss a diffuse flux of neutrinos from the decay of $\phi$ produced in all the past SNe in the universe.
In analogy to the DSNB \cite{Mirizzi:2015eza,Beacom:2010kk,Lunardini:2010ab}, we model the diffuse neutrino flux from the decay of $\phi$ integrated over redshift and weighted by the SN rate,
\begin{align}
    \frac{d\Phi_{\nu_\alpha}}{dE_\nu}=c\int_0^{z_{\rm max}} R_{\rm SN}(z)\frac{dN_{\nu_\alpha}}{dE_\nu'}\left(E_\nu'\right)(1+z)\left|\frac{dt}{dz}\right|dz,
    \label{DSNBphi}
\end{align}
where $z$ is the redshift, $E_\nu'=E_\nu(1+z)$, $|dt/dz|\simeq H_0(1+z)[\Omega_m(1+z)^3+\Omega_\Lambda]^{1/2}$ with the cosmological parameters $H_0\simeq70\ {\rm km\ s^{-1}\ Mpc^{-1}}$, $\Omega_m\simeq0.3$ and $\Omega_\Lambda\simeq0.7$ \cite{Planck:2018vyg}. We take $z_{\rm max}=5$ to be large enough to incorporate almost all SNe events.

The SN rate is proportional to the star-formulation rate $\dot{\rho}^\ast(z)$,
\begin{align}
    R_{\rm SN}(z)=\dot{\rho}^\ast(z)\frac{\int_{8M_{\odot}}^{125M_\odot} dM \psi(M)}{\int_{0.5M_{\odot}}^{125M_\odot} dM M \psi(M)},
    \label{RSN}
\end{align}
where $\psi(M)=dn/dM$ is the initial mass function (IMF). We assume the IMF satisfies a Salpeter law, $\psi(M)\propto M^{-2.35}$.
We use the following fitting formula for the star formulation rate \cite{Yuksel:2008cu,Madau:2014bja,Strolger:2015kra},
\begin{align}
    \dot{\rho}^\ast(z) \propto \left[(1+z)^{p_1k}+\left(\frac{1+z}{5000} \right)^{p_2k}+\left( \frac{1+z}{9} \right)^{p_3k} \right]^{1/k},
\end{align}
where $k=-10,\ p_1=3.4,\ p_2=-0.3$ and $p_3=-3.5$. The SN rate is normalized as $R_{\rm SN}(0)=(1.25\pm 0.5)\times 10^{-4}\ {\rm Mpc}^{-3}\ {\rm yr^{-1}}$ \cite{Lien:2010yb}. Since the uncertainty of the IMF is almost canceled by the ratio in Eq.~(\ref{RSN}), the remaining dominant uncertainty is that of the star formulation rate, where we include this uncertainty in $R_{\rm SN}(0)$.

The quantity $dN_{\nu_\alpha}/dE_\nu'$ is the mean neutrino spectrum induced by the scalar boson decays per one SN. We assume $dN_{\nu_\alpha}/dE_\nu'$ is given by Eqs.~(\ref{phispectrum}) and (\ref{FNfinal}).
In other words, we assume again all the SN core temperature and radius are $T\sim 30\ {\rm MeV}$ and $r_c\sim 10\ {\rm km}$ during a time scale of $t\sim 10\ {\rm s}$, respectively. The chemical potentials for neutrinos are also assumed to be $\mu_{\nu_e}\sim200\ {\rm MeV},\ \mu_{\bar{\nu}_e}\sim-200\ {\rm MeV}$ and $\mu_{\nu_x}=0\ (\nu_x=\nu_\mu,\ \bar{\nu}_\mu,\ \nu_\tau,\ \bar{\nu}_\tau)$ \cite{Burrows:1986me}.
However, strictly speaking, the neutrino spectrum for a SN also depends on the progenitor mass.
In this study we take the simplified model for SNe and leave it as future work a more precise estimation of neutrino spectrum taking into account such details of SNe.

\begin{figure}
	\begin{center}
	\includegraphics[clip,width=10cm]{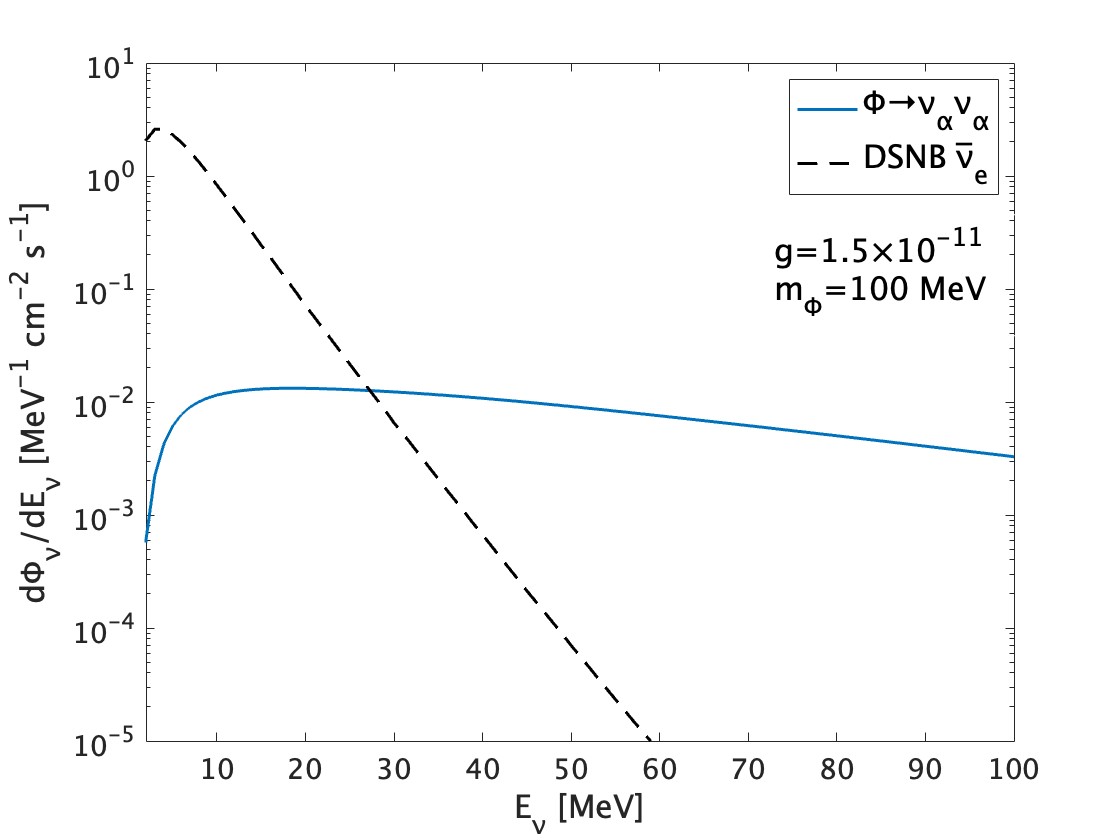}
	\end{center}
	 \vspace{-6mm}
	\caption{The diffuse neutrino flux produced by the decays of the scalar bosons, $\phi\rightarrow \nu_\alpha\nu_\alpha$, emitted from all the past SNe with $g=1.5\times 10^{-11}$ and $m_\phi=100\ {\rm MeV}$ (blue solid line, Eq.~(\ref{DSNBphi})) and the DSNB flux of $\bar{\nu}_e$ directly produced in the SN explosion (black dashed line, Eq.~(\ref{DSNBnue})).}
	\label{fig:DSNBflux_phi}
\end{figure}

In Fig.~\ref{fig:DSNBflux_phi}, we show the diffuse neutrino flux produced by the decays of the scalar bosons, $\phi\rightarrow \nu_\alpha\nu_\alpha$, from all the past SNe with $g=1.5\times 10^{-11}$ and $m_\phi=100\ {\rm MeV}$ (blue solid line), given by Eq.~(\ref{DSNBphi}) and the DSNB flux directly produced in the SN explosion (black dashed line). The details of the DSNB flux of $\bar{\nu}_e$ directly from the SN core (black dashed line) is discussed in Section~\ref{sec6.1.1} and given by Eq.~(\ref{DSNBnue}). The diffuse neutrinos flux produced by the decays of $\phi$ is dominant at $E_\nu\gtrsim 40\ {\rm MeV}$.
Due to the large number density of neutrinos,
the DSNB flux is modified for the small coupling of $g\simeq 10^{-11}$.

\section{Discovery potential for next galactic supernova neutrinos emitted from scalar boson decays}
\label{sec5}
In this section we study the potential of large neutrino detectors to discover the neutrino burst produced by the decay of the light scalar bosons emitted from a future galactic supernova core. 
We will discuss the discovery potential for next galactic supernova neutrinos from vector boson decays in appendix~\ref{appa}.
Here we will assume that a typical distance of a future galactic supernova from the Earth is $d_{\rm SN}=10\ {\rm kpc}$ \cite{Mirizzi:2006xx}.
For neutrino detectors, we will consider Super-Kamiokande (SK) and Hyper-Kamiokande (HK). The SK detector is a water Cherenkov detector with fiducial volume of 22.5 kton in Hida, Japan. The HK experiment is the future experiment of the successor of SK with a larger fiducial volume of 374 kton as the planned final configuration.
In the following we consider fiducial volume of 22.5 kton for SK and 374 kton for HK, respectively.

Although the IceCube detector has the largest volume among neutrino detectors over the world, it is a coarse-grained detector and would obtain only the average energy of the Cherenkov light in the ice, being unable to reconstruct the signal energy \cite{IceCube:2011cwc}. It would also include relatively large background compared with SK and HK  \cite{IceCube:2011cwc}. Other future neutrino experiments, JUNO and DUNE, have smaller fiducial detector volumes than HK. Yet they might have similar discovery potential due to different energy resolutions and cross sections with neutrinos (see also Section \ref{sec6.1}). However, in this section, we will not consider IceCube, JUNO and DUNE.

In the following analysis we will consider only the time-integrated neutrino fluxes and event rates for a conservative estimate of the discovery potential.
\subsection{Event rates at Super-Kamiokande and Hyper-Kamiokande}
\label{sec5.1}
SN neutrinos are mainly captured in SK and HK by the inverse beta decay (IBD) process up to $E_{\nu}\lesssim 80\ {\rm MeV}$ and the quasi-elastic process of neutrinos by the charged current interactions with Oxygen up to $E_\nu \lesssim 1\ {\rm GeV}$ \cite{Super-Kamiokande:2005mbp}. In the following, we only consider the processes involving $\nu_e$ and $\bar{\nu}_e$ for simplicity and a conservative estimate,
\begin{align}
\bar{\nu}_e+p&\rightarrow e^+ + n, \nonumber \\
\overset{(-)}{\nu}_e+ \mathrm{^{16}O}&\rightarrow e^{\pm}+\mathrm{X},
\end{align}
where $\mathrm{X}$ is a nucleon.
The neutrino flux at the Earth from a next galactic SN is obtained from Eq.~(\ref{FNfinal}) multiplied by the geometrical dilution factor $1/(4\pi d_{\rm SN}^2)$.
The number of the expected events is approximately given by
\begin{align}
    \frac{dN}{dE_e}(E_e)&=N_p \frac{1}{4\pi d_{\rm SN}^2}\frac{dN_{\bar{\nu}_e}}{dE_\nu}(E_\nu) \sigma_{\rm IBD}(E_{e},E_\nu) \nonumber \\
    &\ \ \ \ +N_{\mathrm{O}}\left[ \frac{1}{4\pi d_{\rm SN}^2}\frac{dN_{\bar{\nu}_e}}{dE_\nu}(E_\nu) \sigma_{\bar{\nu}_e\textrm{-}\mathrm{O}}(E_{e},E_\nu)
    + \frac{1}{4\pi d_{\rm SN}^2}\frac{dN_{\nu_e}}{dE_\nu}(E_\nu) \sigma_{\nu_e\textrm{-}\mathrm{O}}(E_{e},E_\nu)\right],
    \label{ER1}
\end{align}
where $N_p=1.5\times 10^{33}$ and $N_{\mathrm{O}}=7.5\times 10^{32}$ are the number of free protons and oxygen molecules inside the fiducial volume of 22.5 kton, respectively, and $\sigma_{\rm IBD}$ and $\sigma_{\bar{\nu}_e\textrm{-}\mathrm{O}} (\sigma_{\nu_e\textrm{-}\mathrm{O}})$ are the IBD cross section and the CC cross section between $\bar{\nu}_e$ and $\mathrm{^{16}O}$ ($\nu_e$ and $\mathrm{^{16}O}$), respectively. We take the IBD cross section from Ref.~\cite{Strumia:2003zx} and the CC cross section with $\mathrm{^{16}O}$ from Ref.~\cite{Kolbe:2002gk}. $E_e$ is the energy for electrons and positrons. We have approximately used $E_{e}=E_\nu-1.3\ {\rm MeV}$ for the IBD process, $E_e=E_\nu-11.4\ {\rm MeV}$ for the CC process between $\bar{\nu}_e$ and $\mathrm{^{16}O}$ and $E_e=E_\nu-15.4\ {\rm MeV}$ for the CC process between $\nu_e$ and $\mathrm{^{16}O}$.

Taking into account the energy resolution of the detector, the number of the expected events is then given by
\begin{align}
    \frac{d\tilde{N}}{dE_{e}}=\int_{-\infty}^{\infty} dE_{e}'\frac{1}{\sqrt{2\pi}\delta_E}\exp\left[-\frac{(E_{e}-E_{e}')^2}{2\delta_E^2} \right] \frac{dN}{dE_{e}}(E'_{e}),
    \label{ER2}
\end{align}
where we assume the spectrum is smeared as a Gaussian profile with a energy resolution $\delta_E$. We also assume that the energy resolution of SK and HK is the same as the SK-III observation \cite{Super-Kamiokande:2010tar},
\begin{align}
    \delta_E/E_{e}=0.0349+0.376/\sqrt{(E_{e}/{\rm MeV})}-0.123({\rm MeV}/E_{e}).
    \label{RSK}
\end{align}

The main background is the neutrino events coming directly from a next galactic SN of the standard scenario, which were not produced by the decay of the light bosons.
Since neutrinos produced by the decay of the light bosons from a SN will be observed during a short time of $\sim {\rm max}[10\ {\rm s}, t_D]$, we can neglect other backgrounds. For the background of neutrinos from a SN, we use a “pinched" parametrization of the time-integrated neutrino flux for the different species $\nu=\nu_e, \bar{\nu}_e$ and $\nu_x$ ($\nu_x=\nu_\mu, \nu_\tau, \bar{\nu}_\mu, \bar{\nu}_\tau$) \cite{Keil:2002in,Tamborra:2012ac},
\begin{align}
    \frac{dN_\nu^{\rm BG,0}}{dE}(E)&=L_\nu\frac{(1+\alpha_\nu)^{(1+\alpha_\nu)}}{\Gamma(1+\alpha_\nu)\langle E_\nu \rangle^2}\left(\frac{E}{\langle E_\nu \rangle} \right)^{\alpha_\nu}\exp\left[-(1+\alpha_\nu)\frac{E}{\langle E_\nu \rangle} \right], \nonumber \\
    \alpha_\nu&=\frac{2\langle E_\nu \rangle^2-\langle E_\nu^2 \rangle}{\langle E_\nu^2 \rangle-\langle E_\nu \rangle^2}
    \label{SNnu}
\end{align}
where $L_\nu$ is the luminosity and $\alpha_\nu$ is the so-called pinching parameter. Since the neutrino-sphere radius increases for neutrinos with higher energy and the density of the SN envelope decreases rapidly for the lager radius, the neutrino spectrum become “pinched" as in Eq.~(\ref{SNnu}) , which is numerically confirmed \cite{Keil:2002in,Tamborra:2012ac}. Based on a numerical simulation \cite{Tamborra:2012ac}, we set the values of $L_\nu, \alpha_\nu$ and $\langle E_\nu \rangle$ listed in Table~\ref{tb:ParameterBG}.
During propagation through the SN envelope, flavor conversions are enhanced by their interactions with the electron background called the MSW effect. We neglect flavor conversion effects induced by neutrino-self interactions which are currently under investigation. Considering this effect, the time-integrated flux arriving at the Earth for $\bar{\nu}_e$ is approximately given by \cite{Dighe:1999bi,Lu:2016ipr}, assuming an element of the Pontecorvo-Maki-Nakagawa-Sakata matrix of $|U_{e3}|=0$,
\begin{align}
    \frac{dN_{\nu_e}^{\rm BG,NO}}{dE}(E)&=\frac{dN_{\nu_x}^{\rm BG,0}}{dE},\ \ \ \ 
    \frac{dN_{\bar{\nu}_e}^{\rm BG,NO}}{dE}(E)=c_{12}^2\left(\frac{dN_{\bar{\nu}_e}^{\rm BG,0}}{dE}-\frac{dN_{\nu_x}^{\rm BG,0}}{dE}\right)+\frac{dN_{\nu_x}^{\rm BG,0}}{dE},
    \label{FBG}
\end{align}
where $c_{12}^2=\cos^2\theta_{12}\simeq 0.7$ \cite{Esteban:2020cvm,deSalas:2020pgw}. Here we consider only the normal ordering of neutrino masses for definiteness since the fluxes in both the normal and inverted ordering are almost the same (see e.g., Fig.~175 in Ref.~\cite{Hyper-Kamiokande:2018ofw}). The time-integrated flux of Eq.~(\ref{FBG}) is shown in Fig.~\ref{fig:Nuspectrum_Mdecay} as the black dashed line. Then the number of event for the background can be calculated similarly, following Eqs.~(\ref{ER1}), (\ref{ER2}) and (\ref{FBG}). 

\begin{figure}
	\begin{center}
	\includegraphics[clip,width=10cm]{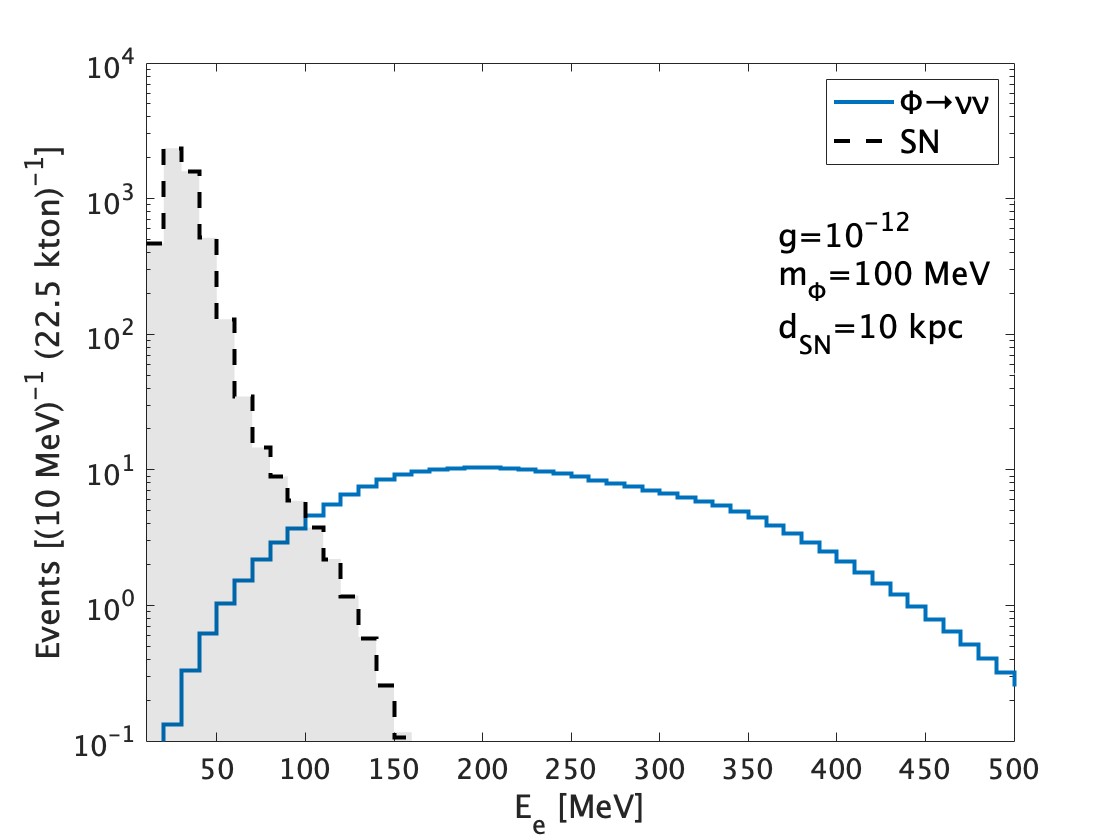}
	\end{center}
	 \vspace{-8mm}
	\caption{Events in Super-Kamiokande with fiducial volume of 22.5 kton from a next galactic SN with $d_{\rm SN}=10\ {\rm kpc}$. We consider $g=10^{-12}, m_{\phi}=100\ {\rm MeV}$. $E_e$ is the energy of electrons and positrons. Blue solid line denotes events from neutrinos produced by the decay of the scalar bosons and black dashed line denotes the background events from the standard SN neutrinos. }
	\label{fig:NextSN_SK_events}
\end{figure}

In Fig.~\ref{fig:NextSN_SK_events}, we show the number of events for neutrinos produced by the decays of the scalar bosons from SN explosion (blue solid line) and neutrinos directly emitted from the supernova (black dashed line). Here we consider $g=10^{-12}$, $m_{\phi}=100\ {\rm MeV}$ and $d_{\rm SN}=10\ {\rm kpc}$. Even for the small coupling of $g=10^{-12}$ with $m_{\phi}=100\ {\rm MeV}$, which is not excluded by current experiments and observations, the signal for $\phi\rightarrow \bar{\nu}_e\bar{\nu}_e$ and $\phi\rightarrow \nu_e\nu_e$ is detectable at $E_{e}\gtrsim 80\ {\rm MeV}$.
The larger cross sections for neutrinos with higher energy help to detect the signal for $\phi\rightarrow \bar{\nu}_e\bar{\nu}_e$ and $\phi\rightarrow \nu_e\nu_e$.

\begin{table}[h]
\begin{center}
	\begin{tabular}{cccc}
		\hline \hline
	     Species & $L_\nu\ ({\rm erg})$ & $\langle E_\nu \rangle\ ({\rm MeV})$ & $\alpha_\nu$   \\
		\hline 
		$\bar{\nu}_e$ & $5\times 10^{52}$ & 13 & 2.8  \\
		$\nu_x$ & $5\times 10^{52}$  & 15 & 2.4 \\
		\hline \hline
	\end{tabular}
	\caption{Parameters for the time-integrated background neutrino fluxes of $\bar{\nu}_e$ and $\nu_x\ (\nu_x=\nu_\mu, \bar{\nu}_\mu, \nu_\tau, \bar{\nu}_\tau)$ from a SN given in Eq.~(\ref{SNnu}) for our reference model, based on Ref.~\cite{Tamborra:2012ac}.}
  \label{tb:ParameterBG}
\end{center}
\end{table}

\subsection{Method and results}
To distinguish SN neutrinos induced by the scalar boson decays from the background, we introduce a cut-off energy $E_{\rm cut}$ defined as \cite{Asai:2022pio}
\begin{align}
    \int_{E_{\rm cut}}^\infty dE_{e}\frac{d\tilde{N}^{\rm BG}}{dE_{e}}=1,
\end{align}
where $d\tilde{N}^{\rm BG}/dE_{e}$ is the background event rate taking into account the energy resolution.
At $E_{e}>E_{\rm cut}$, the background can be considered to be sufficiently small. Then we define the number of the signal,
\begin{align}
    N_{\rm signal}=\int_{E_{\rm cut}}^{\infty}dE_e\frac{d\tilde{N}}{dE_{e}}.
    \label{Nsignal}
\end{align}
We regard the parameter region predicting $N_{\rm signal}= 9$ as the region where we could discover non-standard neutrino interactions with the scalar bosons from a next galactic SN at $99.7\%\ (3\sigma)$ C.L. approximately.

\begin{figure}
	\begin{center}
	\includegraphics[clip,width=10cm]{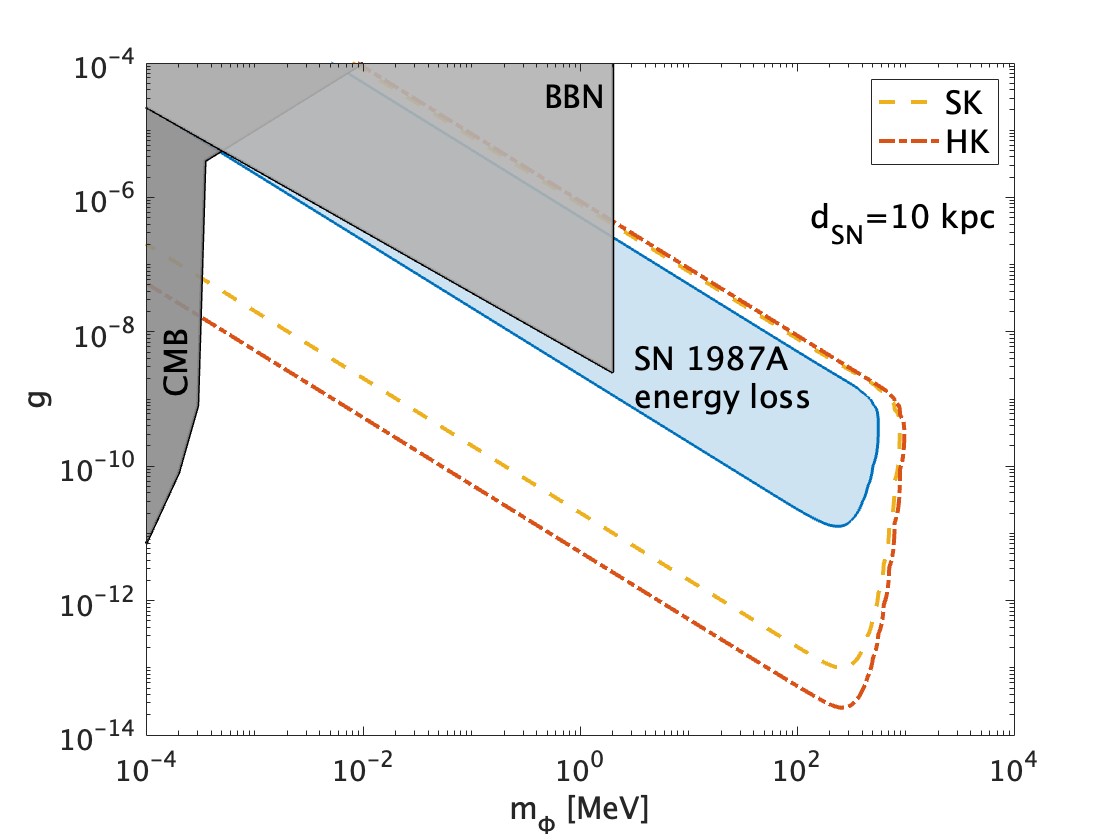}
	\end{center}
	 \vspace{-8mm}
	\caption{Contour plots of $N_{\rm signal}=9$ in the $(m_{\phi},\ g)$ plane for the future observations of next galactic supernova neutrinos produced by the scalar boson decays, $\phi\rightarrow \nu\nu$, in SK (yellow dashed line) and HK (red dot-dashed line) with $d_{\rm SN}=10\ {\rm kpc}$.
	The blue shaded region denotes the current constraints from the energy loss of SN 1987A \cite{Brune:2018sab, Heurtier:2016otg}. The gray shaded regions denote the constraints by the CMB \cite{Archidiacono:2013dua,Escudero:2019gvw} and BBN \cite{Escudero:2019gvw,Huang:2017egl}. We regard $N_{\rm signal}=9$ as $99.7 \%\ (3\sigma)$ C.L. limit on the $(m_{\phi},\ g)$ approximately (see text for details).}
	\label{fig:Const_NextSN}
\end{figure}

In Fig.~\ref{fig:Const_NextSN}, we show the contour plot of $N_{\rm signal}=9$ in the $(m_{\phi},\ g)$ for the future observations of next galactic supernova neutrinos produced by the scalar boson decays in SK and HK with $d_{\rm SN}=10\ {\rm kpc}$.  In this figure, we also show the current excluded region of the $(m_{\phi},\ g)$ by the SN 1987A energy loss\footnote{In SN 1987A observations, we can also constrain non-standard neutrino interactions with $\phi$ from excessive reduction of the trapped lepton number density in the core due to $\nu_e\nu_\alpha \rightarrow \phi$ since the successful bounce shock depends on the ratio between the trapped lepton and baryon number density. However, this constraint strongly depends on the explosion mechanism and numerical simulation. We do not show this deleptonization constraint \cite{Brune:2018sab, Heurtier:2016otg}. } (see the details in Appendix~\ref{appb}) \cite{Brune:2018sab, Heurtier:2016otg}, CMB \cite{Archidiacono:2013dua,Escudero:2019gvw} and BBN \cite{Escudero:2019gvw,Huang:2017egl} observations. Both SK and HK will improve the current constraint on the $(m_{\phi},\ g)$ from SN 1987A significantly. For $m_{\phi}\lesssim 200\ {\rm MeV}$, SK and HK will improve the constraint on the coupling $g$ from SN 1987A by 2-3 orders of magnitude toward the smaller side. For $m_{\phi}\gtrsim 200\ {\rm MeV}$, the discovery potentials in SK and HK will be weaker since the production of the scalar bosons is suppressed by the Boltzmann factor and they are gravitationally trapped. However, since SK and HK are very large detectors, both detectors would still be sensitive to $m_{\phi}\sim 1\ {\rm GeV}$.

In our analysis, we assume the non-standard neutrino interactions with $\phi$ are flavor-universal, $g_{ee}=g_{\mu\mu}=g_{\tau\tau}=g$. Then $\nu_e\nu_e \rightarrow \phi$ is the most dominant production process due to electron capture processes in the SN core. We can neglect the flavor conversion effect in the SN envelope due to the universal couplings. If we consider the non-standard couplings only with $\nu_\mu, \nu_\tau$ and their anti-particle, the sensitivity in SK and HK will be weaker due to the weaker production rate of $\phi$ though the constraints from SN 1987A energy loss also become weaker \cite{Brune:2018sab, Heurtier:2016otg}. For non-universal non-standard neutrino interactions with $\phi$, we will also need to consider the flavor conversion effect in the SN envelope.

\section{Discovery potential for diffuse supernova neutrinos from scalar boson decays}
\label{sec6}
In this section we study the potential of large underground detectors to observe neutrinos produced by the decay of the light scalar bosons from all the past SNe.
We will discuss the discovery potential for diffuse supernova neutrinos from vector boson decays in appendix~\ref{appa}.
Since the flux of such neutrinos is quite small unlike the case of a next galactic SN, we need underground neutrino detector with large volume to distinguish between the signal and relevant backgrounds.
The next-generation neutrino experiments such as Hyper-Kamiokande with Gadolinium (HK with Gd), JUNO and DUNE are expected to be sensitive to this flux. In the following we consider backgrounds and event rates at HK with Gd, JUNO and DUNE and discuss the discovery potential of these experiments for non-standard neutrino interaction with a light boson.
\subsection{Backgrounds and event rates at neutrino detectors}
\label{sec6.1}
First we discuss relevant backgrounds and show event rates for the signal and backgrounds in each neutrino detector.
In this work we compute the number of events for 20 years of data-taking in each experiments.
For calculating the background spectra we have taken help of the publicly available packages \lib{\sng}~\cite{snowglobes} and \lib{nuHawkHunter}~\cite{Barman:2022gjo}.

\subsubsection{Hyper-Kamiokande with Gadolinium}
\label{sec6.1.1}
Hyper-Kamiokande (HK) \cite{Hyper-Kamiokande:2018ofw} is the successor of Super-Kamiokande with a lager water Cherenkov detector to be built in Hida, Japan as we already mentioned in section \ref{sec5}. The planned final configuration consists two tanks, - each of 187 kton of fiducial volume. In the following we consider 374 kton of fiducial volume in total. The main detection channel is the IBD process, $\bar{\nu}_e + p \rightarrow e^+ + n$.

If HK is doped with Gadolinium (Gd), some backgrounds are significantly reduced using neutron tagging \cite{Beacom:2003nk}. A prompt positron event followed by a slightly delayed ($\sim$ tens of $\mu {\rm s}$) emission of $8\ {\rm MeV}\ \gamma$ cascade by excited Gd nucleus (caused by the neutron capture of Gd), is a strong indicator of a true IBD event.
If full neutron tagging is accomplished under doping 0.1 $\%$ Gadolinium to water, we can ignore most of the neutral current (NC) atmospheric neutrinos events which are the main background given by $\gamma$-rays via NC quasi elastic scattering at energies of tens of MeV \cite{T2K:2014vog}, as well as spallation background caused by cosmic rays.
The NC background can further be reduced even more significantly by methods based on Convolution Neutral Network \cite{Maksimovic:2021dmz}.
Only $\mathrm{^9Li}$ spallation products, which emit both positron and neutron, mimicking the IBD process, cannot be neglected but can be substantially reduced at $\gtrsim 12\ {\rm MeV}$ by applying cuts based on a correlation with cosmic ray muons \cite{Hyper-Kamiokande:2018ofw}.  The insivible muon background will be also suppressed by a factor of about 5 using the neutron tagging\cite{Beacom:2003nk}. 
Since we might distinguish the IBD process from the CC $\nu_e$ and $\bar{\nu}_e$ interactions with $\mathrm{^{16}O}$ due to the neutron tagging, we neglect these CC processes, $\overset{(-)}{\nu}_e+ \mathrm{^{16}O}\rightarrow e^{\pm}+\mathrm{X}$ for simplicity.

Then at the positron energy of $E_{e^+}\lesssim10\ {\rm MeV}$, the dominant backgrounds are solar and reactor neutrinos, and $\mathrm{^9Li}$  spallation. At $10\ {\rm MeV}\lesssim E_{e^+} \lesssim 20\ {\rm MeV}$, the dominant background in our study is the DSNB, which is directly produced in the SN explosion \cite{Moller:2018kpn,Tabrizi:2020vmo}, not produced by the decay of the light bosons while at $E_{e^+}\gtrsim 20\ {\rm MeV}$, the main backgrounds are the invisible muon and the charged-current (CC) atmospheric neutrino events. Although the invisible muon events will be suppressed by neutron tagging, they still contribute to the main background at high energies up to $E_{e^+} \sim 50\ {\rm MeV}$, which is the threshold for non-emission of Cherenkov lights induced by muons.

For the calculation of the DSNB events for $\bar{\nu}_e$ directly produced in the SN, we use the following parametrization of the DSNB flux,
\begin{align}
    \frac{d\Phi_{\bar{\nu}_e}^{\rm BG }}{dE_\nu}=c\int^{z_{\rm max}}_0 R_{\rm SN}(z)\frac{dN_{\bar{\nu}_e}^{\rm BG}}{dE_\nu'}(E_\nu')(1+z)\left|\frac{dt}{dz} \right| dz,
    \label{DSNBnue}
\end{align}
where $dN_{\bar{\nu}_e}^{\rm BG}/dE_\nu'$ is given by Eqs.~(\ref{SNnu}) and (\ref{FBG}) with the parameters in Table~\ref{tb:ParameterBG}. We also consider only the normal ordering of neutrino masses for definiteness. The DSNB flux is shown in Fig.~\ref{fig:DSNBflux_phi} as the black dashed line. For recent precise numerical simulations of the DSNB flux, see Refs.~\cite{Horiuchi:2017qja,Kresse:2020nto}.

The expected number of events per energy is approximately, considering the energy resolution of the detector as the Gaussian profile,
\begin{align}
    \frac{d\tilde{N}}{dE_{e^+}}=\epsilon T N_p\int_{\infty}^{\infty}dE'_{e^+} \frac{1}{\sqrt{2\pi}\delta_E}\exp\left[-\frac{(E_{e^+}-E_{e^+}')^2}{2\delta_E^2} \right] \frac{d\Phi_{\bar{\nu}_e}}{dE_\nu}
    \sigma_{\rm IBD}(E'_{e^+},E_\nu),
    \label{ENE}
\end{align}
where $E'_{e^+}=E_\nu-1.3\ {\rm MeV}$, $N_p=2.5\times 10^{34}$ is the number of the proton for 374 kt of fiducial volume and we assume the detector efficiency $\epsilon = 67\%$ (a neutron tagging efficiency of $90\%$ and event selection efficiency of $74\%$) \cite{Hyper-Kamiokande:2018ofw} and $T=20$ years. Here we also assume $\delta_E$ is given by Eq.~(\ref{RSK}).

\begin{figure}
	\begin{center}
	\includegraphics[clip,width=10cm]{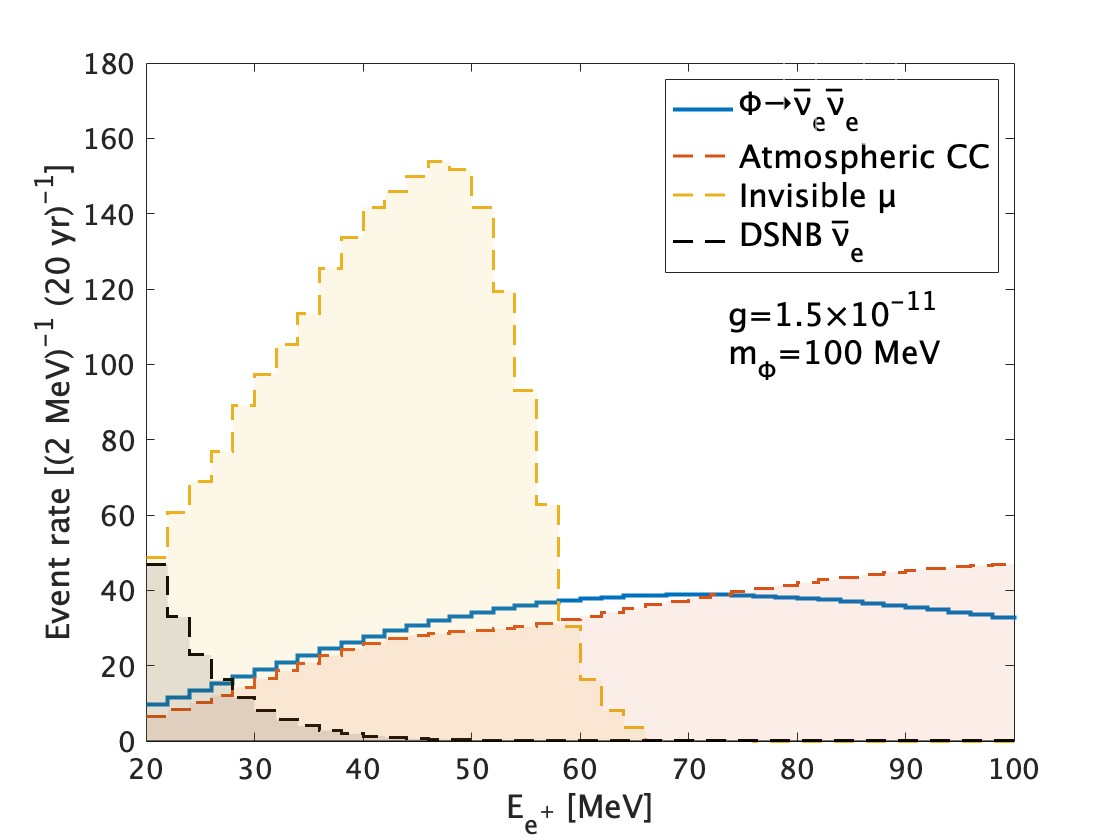}
	\end{center}
	 \vspace{-8mm}
	\caption{Expected event rate predicted for the non-standard neutrino interaction with $g=1.5\times10^{-11}$ and $m_\phi=100\ {\rm MeV}$ as a function of the positron energy in HK with Gd for 20 years of data-taking. The signal due to $\phi\rightarrow \bar{\nu}_e\bar{\nu}_e$ after the emissions of $\phi$ in the SN core is plotted in the blue line. The backgrounds due to the standard DSNB, which is directly produced in the SN, invisible muons and the CC atmospheric neutrino events are plotted in black, yellow, red dashed lines, respectively.}
	\label{fig:DSNB_HK}
\end{figure}

In Fig.~\ref{fig:DSNB_HK}, we show the event rates for the signal due to $\phi\rightarrow \bar{\nu}_e\bar{\nu}_e$ ($g=1.5\times10^{-11}$ and $m_\phi=100\ {\rm MeV}$) after the emission of $\phi$ in the SN core and the backgrounds in HK with Gd in 20 years of data-taking. 
For the invisible muon background between $20\ {\rm MeV} \lesssim E_{e^+} \lesssim 50\ {\rm MeV}$, we use the expected distribution from Fig.~188 of Ref.~\cite{Hyper-Kamiokande:2018ofw} and for this background above $E_{e^+}\gtrsim 50\ {\rm MeV}$, we use the normalized shape from Fig.~11 of Ref.~\cite{Super-Kamiokande:2011lwo}. For estimating atmospheric $\bar{\nu}_e$ background, we use the atmospheric $\bar{\nu}_e$ flux from the result of the Monte Carlo FLUKA simulations \cite{Battistoni:2005pd}.
At $60\ {\rm MeV}\lesssim E_{e^+} \lesssim 100\ {\rm MeV}$, the signal of $\phi\rightarrow \bar{\nu}_e\bar{\nu}_e$ is larger or comparable with the backgrounds. At $E_{e^+}\gtrsim 100\ {\rm MeV}$, the CC atmospheric neutrino events are dominant. The parameter of $g=1.5\times10^{-11}$ and $m_\phi=100\ {\rm MeV}$ is not excluded by the current experiments and observations but is close to the constraints from SN 1987A (see Fig.~\ref{fig:DSNB_discovery} in the next section).

\subsubsection{JUNO}
The Jiangmen Underground Neutrino Observatory (JUNO) \cite{JUNO:2015zny} is a liquid scintillator detector made of linear alkylbenzene ($\mathrm{C_6H_5C_{12}H_{25}}$) with volume of 20 kton to be built in Jiangmen, China. JUNO will have a very good energy resolution because of the liquid scintillator detector. The main detection channel in JUNO is also the IBD process, $\bar{\nu}_e + p \rightarrow e^+ + n$.

Background caused by the neutrinos from nearby nuclear reactors can be safely neglected at $E_{e^+}\gtrsim 10\ {\rm MeV}$. 
The most relevant backgrounds at $E_{e^+}\gtrsim 10\ {\rm MeV}$ are the CC and NC atmospheric neutrino events and the DSNB events. The background events of the fast neutrons produced by muon decays outside the detector happen near the edge of the detector. When we consider a smaller fiducial volume of 17 kton, this background can be reduced. In addition, if pulse-shape discrimination can be successfully applied to the signal and the backgrounds, the fast neutron and NC atmospheric neutrino events can be significantly suppressed \cite{JUNO:2015zny, JUNO:2022lpc}. In the following we apply the pulse shape discrimination to the signal and the backgrounds (see Fig.~3 of Ref.~\cite{JUNO:2022lpc} and Table~5-1 of Ref.~\cite{JUNO:2015zny} for an efficiency of the signal and the backgrounds). At higher energies, the CC atmospheric neutrino events are the dominant background.

The formula of the expected number of events in JUNO is given by Eq.~(\ref{ENE}) with the number of free proton $N_p=1.2\times 10^{33}$, which corresponds fiducial volume of 17 kton, and a detector efficiency for the signal $\epsilon=0.8$ \cite{JUNO:2022lpc}. The width of the Gaussian energy resolution is assumed to be $\delta_E/E_{e^+}=0.03/\sqrt{(E_{e^+}/{\rm MeV})}$ \cite{JUNO:2015zny}.

\begin{figure}
	\begin{center}
	\includegraphics[clip,width=10cm]{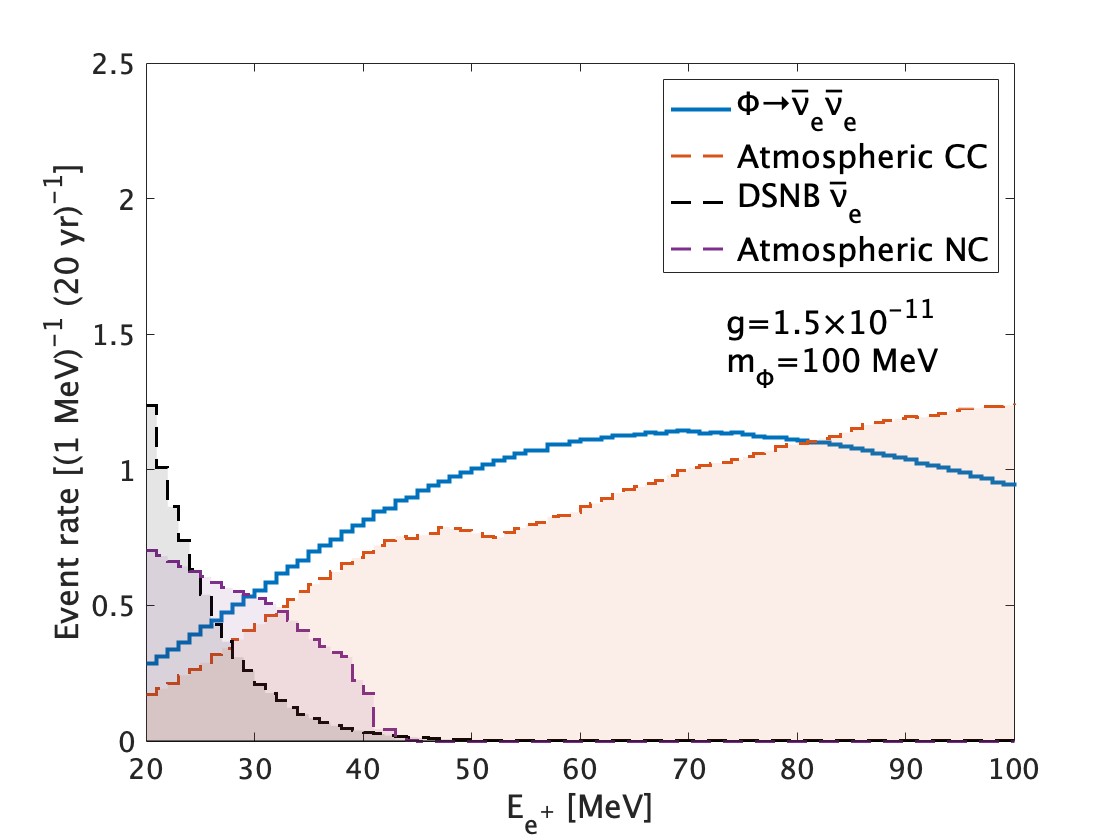}
	\end{center}
	 \vspace{-8mm}
	\caption{Expected event rate predicted for the non-standard neutrino interaction with $g=1.5\times10^{-11}$ and $m_\phi=100\ {\rm MeV}$ as a function of the positron energy in JUNO for 20 years of data-taking.  The signal due to $\phi\rightarrow \bar{\nu}_e\bar{\nu}_e$ after the emissions of $\phi$ in the SN core is plotted in the blue line. The backgrounds due to the standard DSNB, which is directly produced in the SN, the NC and CC atmospheric neutrino events are plotted in black, purple and red dashed lines, respectively.}
	\label{fig:DSNB_JUNO}
\end{figure}

In Fig.~\ref{fig:DSNB_JUNO}, we show the event rates for the signal due to $\phi\rightarrow \bar{\nu}_e\bar{\nu}_e$ ($g=1.5\times10^{-11}$ and $m_\phi=100\ {\rm MeV}$) after the emission of $\phi$ in the SN core and the backgrounds in JUNO in 20 years of data-taking.
For the NC atmospheric neutrino background, we make use the event rate from the right panel of Fig.~5-2 of Ref.~\cite{JUNO:2015zny}.
For CC atmospheric $\bar{\nu}_e$ background, we also use the atmospheric $\bar{\nu}_e$ flux from the result of the Monte Carlo FLUKA simulations \cite{Battistoni:2005pd}. Since the location of JUNO is almost the same latitude as that of SK but their flux is slightly smaller for $E_\nu\sim 50\ {\rm MeV}$ \cite{Wurm:2007cy}, we use to rescale Table~3 for SK of Ref.~\cite{Battistoni:2005pd} by a factor of 0.9.
At $40\ {\rm MeV}\lesssim E_{e^+} \lesssim 100\ {\rm MeV}$, the signal of $\phi\rightarrow \bar{\nu}_e\bar{\nu}_e$ is larger or comparable with the backgrounds. At $E_{e^+}\gtrsim 100\ {\rm MeV}$, the CC atmospheric neutrino events are dominant.

\subsubsection{DUNE}
The Deep Underground Neutrino Experiment (DUNE) \cite{DUNE:2020ypp} is a liquid argon detector with fiducial volume of 40 kton to be built in South Dakota, USA. The main detection channel is the CC interaction of $\nu_e$ with argon,
\begin{align}
    \nu_e +\mathrm{^{40}Ar} \rightarrow e^-+\mathrm{^{40} K^\ast}.
\end{align}

The details of the background in DUNE are still under study. In the following, we assume the background will be similar to the one in ICARUS detector \cite{Cocco:2004ac}.
The dominant backgrounds at $E_{\nu}\lesssim 19\ {\rm MeV}$ are the $\mathrm{^8B}$ and hep solar neutrino events while at $E_{\nu}\gtrsim 19\ {\rm MeV}$, the relevant background is the DSNB events. The CC atmospheric $\nu_e$ event is also another dominant at $E_\nu\gtrsim 30\ {\rm MeV}$.
We calculate the DSNB events for $\nu_e$ similarly to the calculation of the DSNB events for $\bar{\nu}_e$ discussed in section \ref{sec6.1.1}.

The expected number of events per neutrino energy in DUNE is, considering the energy resolution as the Gaussian profile,
\begin{align}
     \frac{d\tilde{N}}{dE_{\nu}}=\epsilon T N_{\rm Ar} \int_{\infty}^{-\infty} dE_\nu' \frac{1}{\sqrt{2\pi}\delta_E}\exp\left[-\frac{(E_\nu-E_\nu')^2}{2\delta_E^2} \right] \sigma_{\nu-{\mathrm{Ar}}}(E_\nu') \frac{d\Phi_{\nu_e}}{dE_\nu'},
\end{align}
where $N_{\rm Ar}=6.02\times 10^{32}$ is the number of argon in 40 kton liquid argon and we assume a detector efficiency of $\epsilon=86 \%$ (a trigger efficiency of $90 \%$ and a reconstruction efficiency of $96 \%$). $\sigma_{\nu-{\rm Ar}}$ is the total cross section between $\nu_e$ and $\mathrm{Ar}$ and we take this cross section from Ref.~\cite{GilBotella:2003sz}. The resolution for neutrino energy in the DUNE detector is assumed to be $\delta_E/E_\nu=0.2$ (see Fig.~7.5 of Ref.~\cite{DUNE:2020ypp}). 

\begin{figure}
	\begin{center}
	\includegraphics[clip,width=10cm]{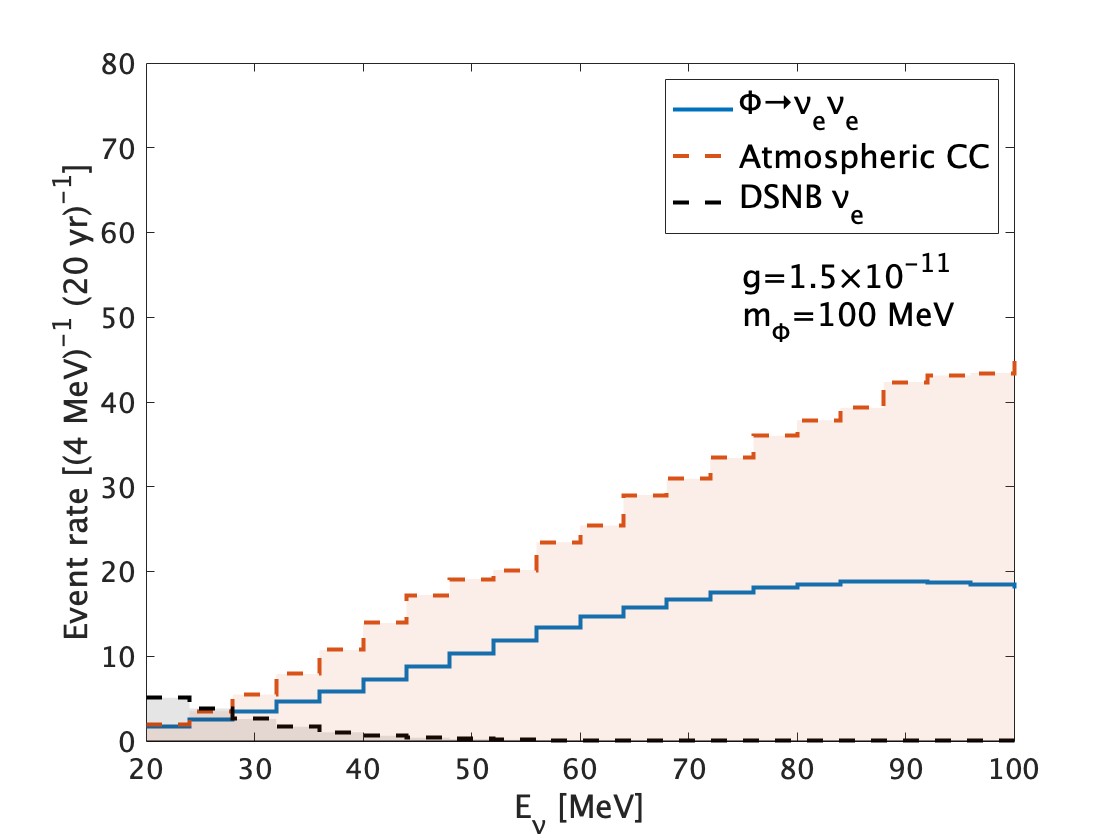}
	\end{center}
	 \vspace{-8mm}
	\caption{Expected event rate predicted for the non-standard neutrino interaction with $g=1.5\times10^{-11}$ and $m_\phi=100\ {\rm MeV}$ as a function of the neutrino energy in DUNE for 20 years of data-taking. The signal due to $\phi\rightarrow \nu_e\nu_e$ after the emissions of $\phi$ in the SN core is plotted in the blue line. The backgrounds due to the standard DSNB, which is directly produced in the SN, the CC atmospheric neutrino events are plotted in black red dashed lines, respectively.}
	\label{fig:DSNB_DUNE}
\end{figure}

In Fig.~\ref{fig:DSNB_DUNE}, we show the event rates for the signal due to $\phi\rightarrow \nu_e\nu_e$ ($g=1.5\times10^{-11}$ after the emissions in the SN core and $m_\phi=100\ {\rm MeV}$) and the backgrounds in the DUNE in 20 years of data-taking. 
For atmospheric $\bar{\nu}_e$ background, we use the atmospheric $\nu_e$ flux from the result of the Monte Carlo FLUKA simulations \cite{Battistoni:2005pd}. Since the location of DUNE is almost the same latitude as that of the Gran-Sasso laboratory, we use Table~5 of Ref.~\cite{Battistoni:2005pd}.
Unfortunately the CC atmospheric neutrino background are dominant compared with the signal in all energy region. Since in the DUNE location the atmospheric neutrino flux is larger than in the locations for HK and JUNO \cite{Battistoni:2005pd}, DUNE has weaker sensitivity to the signal than HK with Gd and JUNO.

\subsection{Method and results}
To estimate sensitivities of the non-standard neutrino interaction with the scalar boson, we perform a $\chi^2$ analysis for each detector considering the systematic uncertainties using a pull method \cite{Fogli:2002pt}. In all cases, we assume 20 years of data-taking and an energy bin of 2 MeV for HK with Gd, 1 MeV for JUNO and 4 MeV for DUNE. The energy range for HK with Gd, JUNO and DUNE is assumed to be 60-100 MeV, 40-100 MeV and 40-100 MeV, respectively. Since the standard DSNB directly produced in the SN explosion might include large uncertainty due to the emission of $\phi$, we adopt the above energy range, neglecting the small event due to the standard DSNB.
We define the $\chi^2$ function as \cite{Moller:2018kpn}
\begin{align}
    \chi^2=\min_{a,b,c}\left[-2\sum_i\ln\frac{L_{0,i}}{L_{1,i}}+\frac{a^2}{\sigma_a^2}+\frac{b^2}{\sigma_b^2}+\frac{c^2}{\sigma_c^2} \right],
    \label{chi2}
\end{align}
where the summation is over the energy bins relevant to the experiment. We consider the Poisson likelihood functions as
\begin{align}
    L_{x,i}=\frac{\lambda_{x,i}^{k_i}\exp(-\lambda_{x,i})}{k_i!}
\end{align}
with $x=0,1$. Thus $L_{x,i}$ describes the probability to observe $k_i$ events at the $i$-th energy bin with a Poisson distribution with mean $\lambda_{x,i}$. We take 
\begin{align}
    \lambda_{0,i}=(1+b)N_{{\rm BG},i},\ \ \ \ \lambda_{1,i}=k_i=(1+a)N_{{\rm signal},i}+(1+b)N_{{\rm BG},i}.
\end{align}
Only in the case of DUNE, we consider $\lambda_{0,i}=(1+b)(1+c)N_{{\rm BG},i}$ and $\lambda_{1,i}=k_i=(1+a)(1+c)N_i+(1+b)(1+c)N_{{\rm BG},i}$.
The pull parameters $a,b$ and $c$ take into account uncertainties for the diffuse SN flux, the background and the neutrino-argon cross section respectively.
The uncertainty for the diffuse SN flux is mainly the SN rate and we take the standard deviation to be $\sigma_a=30 \%$.
The main uncertainty for the background is the one for the atmospheric neutrino flux prediction in the low energy ($\lesssim 1 {\rm GeV}$) and we assume $\sigma_b=25\%$ \cite{Battistoni:2005pd}. The standard deviation for the neutrino-algon cross section is taken to be $\sigma_c=15\%$ \cite{ArgoNeuT:2014rlj}. The set of the pull-parameters $a,b$ and $c$ are marginalized in Eq.~(\ref{chi2}).

\begin{figure}
	\begin{center}
	\includegraphics[clip,width=10cm]{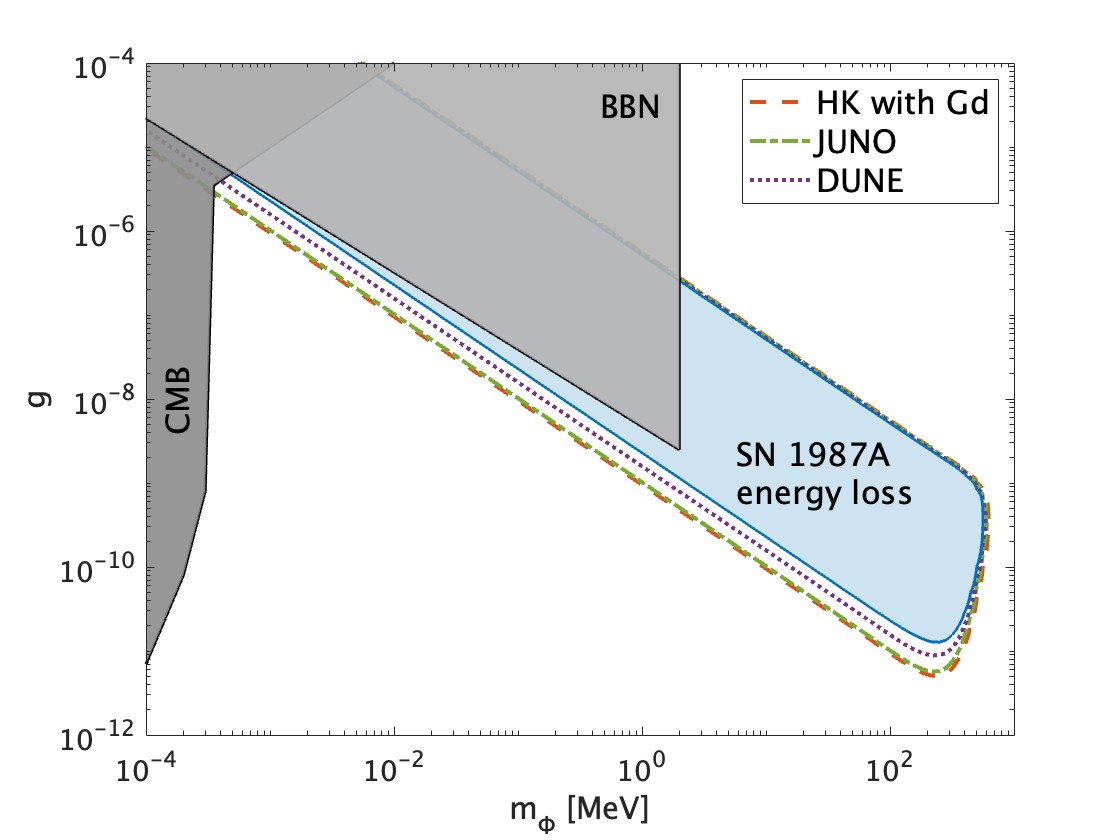}
	\end{center}
	 \vspace{-8mm}
	\caption{Contour plots of $90\ \%$ C.L. intervals in the $(m_{\phi},\ g)$ plane for the future observations of neutrinos produced by the scalar boson decays from all the past CC-SNe, $\phi\rightarrow \nu\nu$, in HK with Gd (red dashed line), JUNO (green dot-dashed line) and DUNE (purple dotted line).
	The blue shaded region denotes the current constraints from the energy loss of SN 1987A \cite{Brune:2018sab, Heurtier:2016otg}. The gray shaded regions denote the constraints by the CMB \cite{Archidiacono:2013dua,Escudero:2019gvw} and BBN\cite{Escudero:2019gvw,Huang:2017egl}.}
	\label{fig:DSNB_discovery}
\end{figure}

In Fig.~\ref{fig:DSNB_discovery}, we show $90\ \%$ C.L. ($\chi^2=2.71$) intervals in the $(m_{\phi},\ g)$ plane for the future observations of neutrinos produced by the scalar boson decays from all the past SNe, $\phi\rightarrow \nu\nu$, in HK with Gd (red dashed line), JUNO (green dot-dashed line) and DUNE (purple dotted line). We also show the current excluded region of the $(m_{\phi},\ g)$ by the SN 1987A energy loss (see the details in Appendix~\ref{appb}) \cite{Brune:2018sab, Heurtier:2016otg}, CMB \cite{Archidiacono:2013dua,Escudero:2019gvw} and BBN \cite{Escudero:2019gvw,Huang:2017egl}. All of HK with Gd, JUNO and DUNE will slightly improve the constraint from SN 1987A. HK with Gd and JUNO have the similar sensitivity, improving the constraint from SN 1987A by a factor of $\sim 2$. 
In terms of the total emitted energy of the scalar bosons $\phi$, this improvement corresponds $25\%$ of the total energy of SN 1987A since the emission rate of $E_\phi$ is proportional to $g^2$.
Though the expected observations of the diffuse supernova neutrinos produced by the decays will improve less the current constraints on the $(m_{\phi},\ g)$, it will be also useful in complementing the constraints from the energy loss of SN 1987A \cite{Brune:2018sab, Heurtier:2016otg}.

We should note that we have assumed the core in all the past SNe with the same radius of $r_c\sim 10\ {\rm km}$, the same temperature of $T\sim 30\ {\rm MeV}$ and the same chemical potential for $\nu_e$ and $\bar{\nu}_e$ of $\mu_{\nu_e}\sim-\mu_{\bar{\nu}_e}\sim-200\ {\rm MeV}$ during a time scale of $t\sim 10\ {\rm s}$. Strictly, the profile of the core and the resultant spectrum of $\phi$ would depend on the progenitor mass of the SN. Thus, this assumption might include uncertainty than the improvement in our analysis of a factor of $\sim 2$. However, it will still complement the constraint from the energy loss of SN 1987A.

\section{Conclusions}
\label{sec7}
We have investigated the potential of future observations of next galactic supernova neutrinos and neutrinos from all the past SNe to probe flavor-universal non-standard neutrino interactions, $g_{ee}=g_{\mu\mu}=g_{\tau\tau}=g$, with a light massive boson. In Sections~\ref{sec3} and \ref{sec4}, we have shown the massive scalar boson produced in the SN core can decay back into neutrinos, $\phi\rightarrow \nu\nu$ , in the steller envelope, modifying the SN neutrino flux without violating the constraint from the energy loss of SN 1987A.
Our main findings are:

\begin{itemize}
  \setlength{\parskip}{0mm}
  \setlength{\itemsep}{1mm}
  \item

The discovery potential of SK and HK to probe neutrino burst produced by the decays of the light scalar bosons from a next galactic supernova at a typical distance of $d_{\rm SN}=10\ {\rm kpc}$ is shown in Fig.~\ref{fig:Const_NextSN}.
Then the sensitivities of SK and HK to the neutrino coupling with the scalar boson are up to $g\lesssim 10^{-13}$ and $g\lesssim 2\times 10^{-14}$, respectively, exceeding the constraint from the energy loss of SN 1987A by several orders of magnitude. Due to the large statistics of neutrino events, SK and HK would be still sensitive to $m_{\phi}\sim 1\ {\rm GeV}$ where the production of the scalar bosons in the core are suppressed by the boltzmann factor $e^{-m_\phi/T}$ and the gravitational trapping of the core.

\item
The discovery potential of HK with Gd, JUNO and DUNE to probe the neutrino burst produced by the decays of the light scalar bosons from all the past SNe is shown in Fig.~\ref{fig:DSNB_discovery}.
Their sensitivities to the neutrino coupling with the scalar boson slightly improve the constraint from the energy loss of SN 1987A by a factor of $\sim 2$. However, we have assumed a simple supernova core model for all the past SNe with the core temperature $T\sim 30\ {\rm MeV}$, the core radius $r_c\sim 10\ {\rm km}$, the chemical potential for the (anti-)electron neutrino $\mu_{\nu_e}\sim -\mu_{\bar{\nu}_e}\sim 200\ {\rm MeV}$, and the burst duration time $\Delta t = 10\ {\rm s}$. This assumption might include an uncertainty larger than the improvement in our analysis of a factor of $\sim 2$.

\end{itemize}

For the flavor-universal neutrino couplings to a light scalar boson as in our analysis, the light bosons are dominantly produced by the process $\nu_e\nu_e\rightarrow \phi$ in the SN core. On the other hand, if one considers flavor-dependent neutrino couplings with $\phi$,
both the discovery potential for $\phi$ and the constraint from the energy loss of SN1987A are to be changed, because the production rate of $\phi$ in the SN core and neutrino flavor conversions in the stellar envelope will be different.
We have also assumed the SN core with $T\sim 30\ {\rm MeV},\ r_c\sim 10\ {\rm km}$ and $\mu_{\nu_e}\sim -\mu_{\bar{\nu}_e}\sim 200\ {\rm MeV}$ during $\Delta t = 10\ {\rm s}$. More improved simulations of SN explosion, including the emission of $\phi$, would be interesting future work.

In appendix \ref{appa}, we also estimate the discovery potentials for next galactic supernova and diffuse supernova neutrinos from vector boson decays with flavor-universal couplings to neutrinos.
We found that the future observations of next galactic and diffuse supernova neutrinos can be sensitive to the couplings, improving the limit from the SN 1987A energy loss by several orders of magnitude and a factor of $\sim 2$, respectively while the sensitivities (and the constraint of SN 1987A energy loss) for the vector boson are weaker than those for the scalar boson.

To summarize, future neutrino observations from a next galactic SN will significantly improve the sensitivity to non-standard neutrino interactions with a light boson beyond the corresponding constraint from the energy loss argument for SN 1987A, and future neutrino observations from all the past SNe will complement the SN1987A limit.


\section*{Acknowledgments}
We thank the anonymous referee for helpful comments. This work was supported by IBS under the project code, IBS-R018-D1.


\appendix

\section{Discovery potentials for supernova neutrinos from vector boson decays}
\label{appa}

In this appendix we consider flavor-universal non-standard neutrino interactions with a light vector boson.
Then we estimate the potential of neutrino detectors to discover next galactic supernova and diffuse supernova neutrinos produced by the decay of the vector bosons emitted from the SN core. 

Flavor-universal non-standard neutrino interactions with a vector boson may be described by
\begin{align}
\mathcal{L}=g_{\alpha\beta} \bar{\nu}_\alpha  Z_\mu'\gamma^\mu P_L \nu_\beta,
\end{align}
where $P_L=(1-\gamma_5)/2$ is the left chirality projection operator and we assume
\begin{align}
g_{ee}=g_{\mu\mu}=g_{\tau\tau}=g,
\end{align}
and the other components are zero. The dominant production processes for $Z'$ are $\nu_\alpha \bar{\nu}_\alpha \rightarrow Z'$ and subsequently the decay processes $Z' \rightarrow \nu_\alpha \bar{\nu}_\alpha$ occur for a massive vector boson.
The squared matrix amplitude for $Z' \rightarrow \nu_\alpha \bar{\nu}_\alpha$ and the decay rate in the rest frame of $Z'$ are given by
\begin{align}
|\mathcal{M}|^2_{Z' \rightarrow \nu_\alpha \bar{\nu}_\alpha}=\frac{2g^2}{3} m_{Z'}^2,\ \ \ \ \ \ \ \  \Gamma_{Z' \rightarrow \nu_\alpha \bar{\nu}_\alpha}=\frac{g^2}{24\pi}m_{Z'}.
\end{align}

In a way similar to the case of the scalar boson discussed in sections~\ref{sec3}, \ref{sec4}, \ref{sec5} and \ref{sec6}, we can estimate the discovery potential for supernova neutrinos from the vector boson decays.
The dominant production process for the vector boson $Z'$ in the SN core is $\nu_\alpha\bar{\nu}_\alpha \rightarrow Z'$ while that for the scalar boson $\phi$ is $\nu_e\nu_e\rightarrow \phi$ since the electron neutrinos are most populated in the core by the electron capture process. Then the sensitivities of neutrino detectors for supernova neutrinos produced by the scalar and vector bosons will be quite different.

\begin{figure}
	\begin{center}
	\includegraphics[clip,width=10cm]{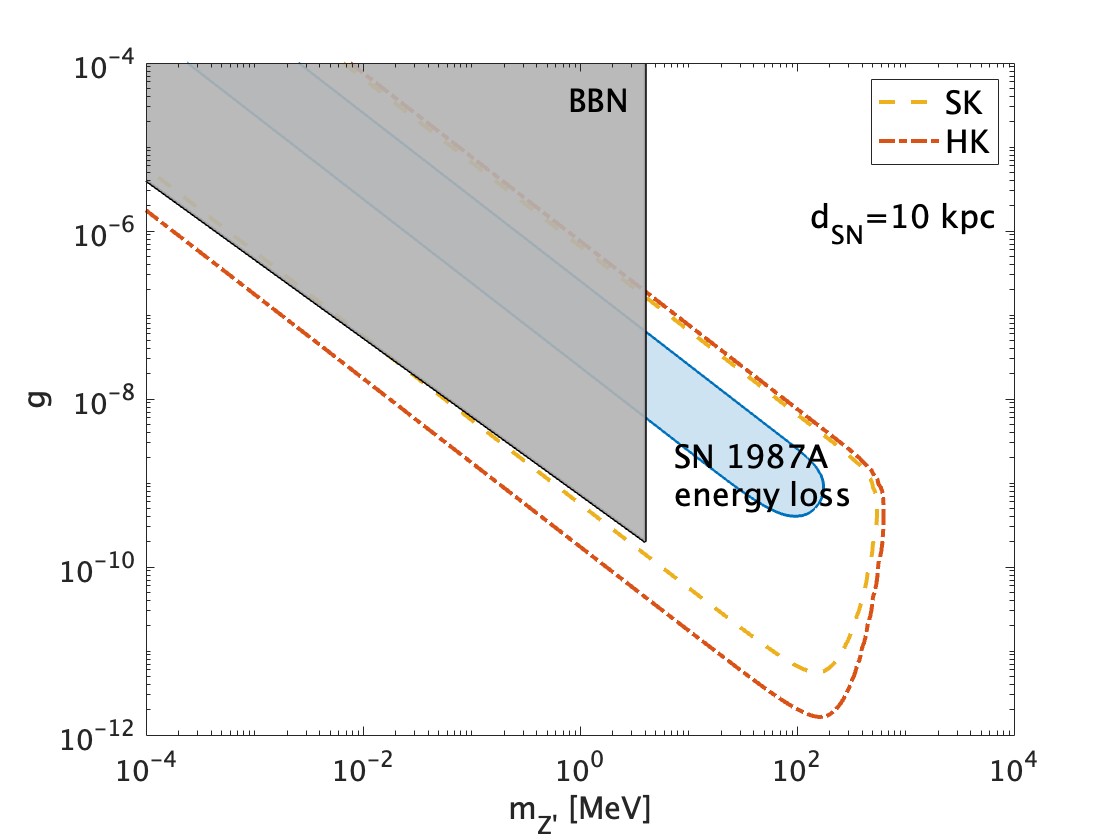}
	\end{center}
	 \vspace{-8mm}
	\caption{Contour plots of $N_{\rm signal}=9$ in the $(m_{Z'},\ g)$ plane for the observations of next galactic supernova neutrinos produced by the vector boson decays, $Z'\rightarrow \nu\bar{\nu}$, in SK (yellow dashed line) and HK (red dot-dashed line) with $d_{\rm SN}=10\ {\rm kpc}$.
	The blue shaded region denotes the current constraints from the energy loss of SN 1987A \cite{Brune:2018sab, Heurtier:2016otg}. The gray shaded regions denote the constraints by BBN \cite{Huang:2017egl}. We regard $N_{\rm signal}=9$ as $99.7 \%\ (3\sigma)$ C.L. limit on the $(m_{Z'},\ g)$ approximately (see text in section~\ref{sec5.1} for details).}
	\label{fig:Const_NextSN_vector}
\end{figure}

In Fig.~\ref{fig:Const_NextSN_vector}, we show the contour plot of $N_{\rm signal}=9$ defined in Eq.~(\ref{Nsignal}) in the $(m_{Z'},\ g)$ for the future observations of next galactic supernova neutrinos from the vector boson decyas in SK and HK with a distance between the supernova and the Earth of $d_{\rm SN}=10\ {\rm kpc}$.
We also show the current excluded region of the  $(m_{Z'},\ g)$ by the SN 1987A energy loss~\cite{Brune:2018sab, Heurtier:2016otg} and BBN observations~\cite{Huang:2017egl}.
Both SK and HK will significantly improve the current constraint on the $(m_{Z'},\ g)$ from SN 1987A even in the case of vector boson. Compared with the case of the scalar boson, the sensitivity for the vector boson will be weaker as expected above. 

\begin{figure}
	\begin{center}
	\includegraphics[clip,width=10cm]{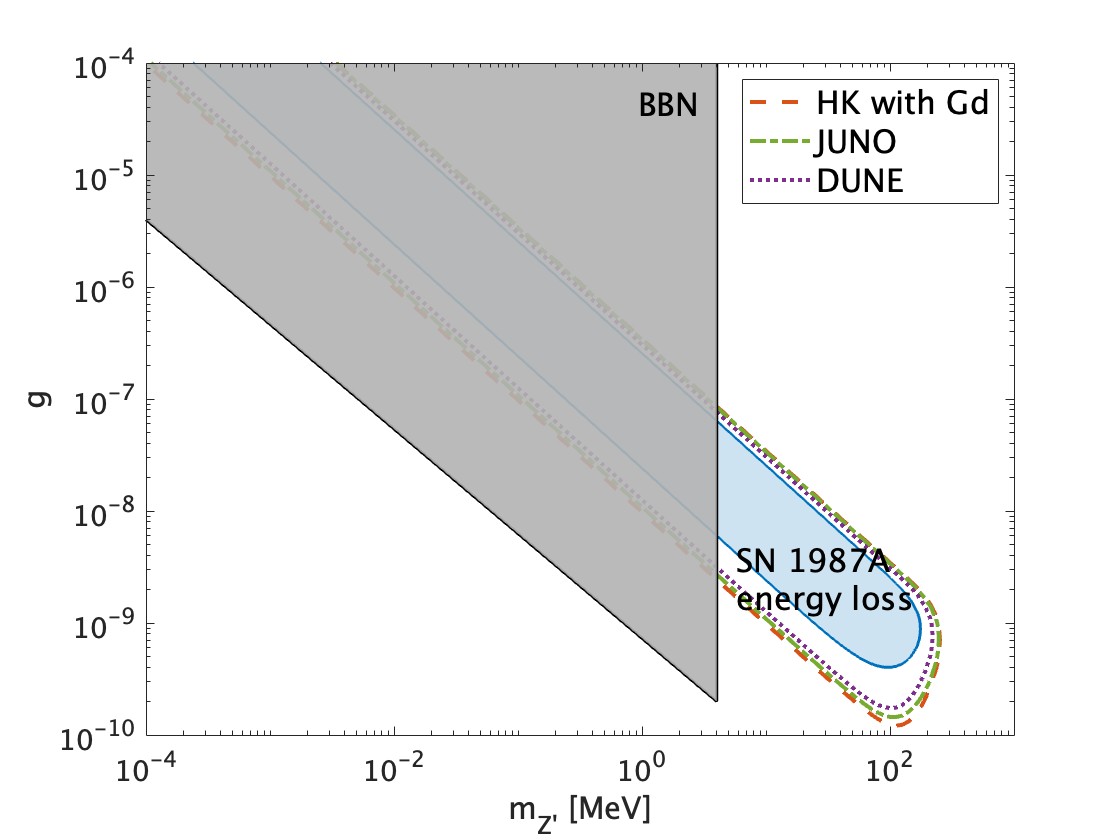}
	\end{center}
	 \vspace{-8mm}
	\caption{Contour plots of $90\ \%$ C.L. intervals in the $(m_{Z'},\ g)$ plane for the future observations of neutrinos produced by the vector boson decays from all the past CC-SNe, $Z'\rightarrow \nu\bar{\nu}$, in HK with Gd (red dashed line), JUNO (green dot-dashed line) and DUNE (purple dotted line).
	The blue shaded region denotes the current constraints from the energy loss of SN 1987A \cite{Brune:2018sab, Heurtier:2016otg}. The gray shaded regions denote the constraints by BBN\cite{Huang:2017egl}.}
	\label{fig:DSNB_discovery_vector}
\end{figure}

In Fig.~\ref{fig:DSNB_discovery_vector}, we show $90\ \%$ C.L. intervals in the $(m_{Z'},\ g)$ plane for future observations of neutrinos produced by the vector boson decays from all the past SNe, i.e. the DSNB, in HK with Gd (red dashed line), JUNO (green dot-dashed line) and DUNE (purple dotted line) with 20 years of data-taking.
All of HK with Gd, JUNO and DUNE will improve the constraint from SN 1987A by a factor of $\sim 2$.
However, these result might also be subject to a large uncertainty concerning the precise estimation of the diffuse supernova neutrino flux.

\section{Constraint from the energy loss of SN 1987A}
\label{appb}

In this appendix we discuss the details of constraint on non-standard neutrino interactions with the massive scalar boson in Eqs.~(\ref{NNI}) and (\ref{NNI2}) from excessive energy loss of SN 1987A neutrino burst, following Refs.~\cite{Brune:2018sab, Heurtier:2016otg}. In Refs.~\cite{Brune:2018sab, Heurtier:2016otg}, flavor-dependent non-standard neutrino interactions are discussed. In this study, we have reproduced the constraint on the flavor-universal non-standard neutrino interactions in Eqs.~(\ref{NNI}) and (\ref{NNI2}).

The energy loss rate per volume and time due to $\nu_\alpha(p_1)\nu_\alpha(p_2)\rightarrow \phi(p_\phi)$ and $\bar{\nu}_\alpha(p_1)\bar{\nu}_\alpha(p_2)\rightarrow \phi(p_\phi)$ in the SN core is
\begin{align}
    \frac{d\rho_\phi}{dt}=\sum_\nu\int d\Pi_\phi d\Pi_1d\Pi_2 E_\phi S|\mathcal{M}|^2_{\nu\nu\rightarrow\phi}(2\pi)^4\delta^4(p_1+p_2-p_\phi)f_\nu(p_1)f_\nu(p_2).
\end{align}
The difference with the production rate of $\phi$ in Eq.~(\ref{Bphi}) is only a factor of $E_\phi$. We consider again the SN core is an isotropic and homogeneous with $r_c\sim10\ {\rm km}$ and $T\sim 30\ {\rm MeV}$ during a time scale of $\Delta t=10\ {\rm s}$ for simplicity. Neutrino chemical potentials in the core are also assumed to be $\mu_{\nu_e}\sim -\mu_{\bar{\nu}_e} \sim 200\ {\rm MeV}$ and $\mu_{\nu_x}\simeq 0\ (\nu_x=\nu_\mu,\bar{\nu}_\mu,\nu_\tau,\bar{\nu}_\tau)$ during $\Delta t$. In the core, we consider neutrinos follow the Fermi-Dirac distribution in thermal equilibrium. Then the total energy loss is, including the survival probability of $\phi$ outside the neutrino-sphere with $r_\nu\sim 50\ {\rm km}$ and the gravitational trapping to the SN core discussed in Section~\ref{sec3},
\begin{align}
    E_{\phi}&=V\Delta t\sum_\nu\int d\Pi_\phi d\Pi_1d\Pi_2 E_\phi S|\mathcal{M}|^2_{\nu\nu\rightarrow\phi}(2\pi)^4\delta^4(p_1+p_2-p_\phi)f_\nu(p_1)f_\nu(p_2) \nonumber \\
    &\ \ \ \ \ \ \ \  \times e^{-\Gamma_\phi r_\nu/\gamma}\chi(E_\phi-K_{\rm tr}-m_\phi),
\end{align}
with $V=\frac{4}{3}\pi r_c^3$, $\gamma=E_\phi/m_\phi$. $\chi$ is the the Heaviside step function and $K_{\rm tr}$ is given by Eq.~(\ref{Potentialenergy}).
Due to the fact that we observed the neutrino burst from SN 1987A with the total energy of the SN explosion of $\sim 3\times 10^{53}\ {\rm erg}$, we need to avoid excessive energy loss of SN 1987A neutrino burst. Then we demand
\begin{align}
    E_\phi \lesssim 3\times 10^{53}\ {\rm erg}.
\end{align}

\bibliographystyle{JHEP}
\bibliography{reference}

\end{document}